\documentclass[onecolumn]{aastex63}
\usepackage{xcolor}
\graphicspath{{./}{figures/}}
\newcommand{\be}{\begin{equation}}
\newcommand{\ee}{\end{equation}}
\newcommand{\ba}{\begin{eqnarray}}
\newcommand{\ea}{\end{eqnarray}}
\newcommand{\bnabla}{\mbox{\boldmath$\nabla$}}

\newcommand{\bbeta}{\mbox{\boldmath$\beta$}}

\newcommand{\nn}{\mbox{} \nonumber \\ \mbox{}}

\usepackage{mathtools}

\begin{document}

\title{Radio Emission of Pulsars. II. \\ Coherence Catalyzed by Cerenkov-Unstable Shear Alfv\'en Waves}

\author{Christopher Thompson}
\affiliation{Canadian Institute for Theoretical Astrophysics, 60 St. George Street, Toronto, ON M5S 3H8, Canada}  

\shortauthors{Thompson}
\shorttitle{Radio Emission of Pulsars. II. Cerenkov-Unstable Shear Alfv\'en Waves}
\accepted{4 February 2022}

\begin{abstract}
  This paper explores small-scale departures from force-free electrodynamics around a rotating neutron star,
  extending our treatment of resistive instability in a quantizing magnetic field.
  A secondary, Cerenkov instability is identified:  relativistic particles flowing through
  thin current sheets excite propagating charge perturbations that are localized
  near the sheets.  Growth is rapid at wavenumbers below the inverse ambient skin depth $k_{p,\rm ex}$.
  Small-scale Alfv\'enic wavepackets are promising sources of coherent curvature radiation.  When the group Lorentz factor
  $\gamma_{\rm gr} \lesssim (k_{p,\rm ex}R_c)^{1/3} \sim 100$, where $R_c$ is the magnetic curvature radius,
  a fraction $\sim 10^{-3}$-$10^{-2}$ of the particle kinetic energy is radiated into the extraordinary mode
  at a peak frequency $\sim 10^{-2}ck_{p,\rm ex}$.  Consistency with
  observations requires a high pair multiplicity ($\sim 10^{3-5}$) in the pulsar magnetosphere.
  Neither the primary, slow resistive instability nor the secondary, Alfv\'enic instability depend directly on the presence 
  of magnetospheric `gaps', and may activate where the mean current is fully supplied by outward drift of the corotation
  charge.    The resistive mode is overstable and grows at a rate comparable to the stellar spin frequency;
  the model directly accommodates strong pulse-to-pulse radio flux
  variations and coordinated sub-pulse drift.  Alfv\'en mode growth can track the local plasma conditions,
  allowing for lower-frequency emission from the outer magnetosphere.  Beamed radio emission from charged packets with
  $\gamma_{\rm gr} \sim 50-100$  also varies on sub-millisecond timescales.  The modes identified here will be
  excited inside the magnetosphere of a magnetar, and may mediate Taylor relaxation of the magnetic twist.
  \end{abstract}


\section{Introduction}\label{s:intro}

Theoretical approaches to pulsar radio emission have a curious history.   Very quickly a simple physical picture
predicated on charge clumping in an electromagnetic cascade was developed, giving a transparent explanation for the observed
linear polarization swings
\citep{rc69,rs75}.  But over five decades, no consensus has emerged on how the required charged soliton structures would develop;
theorists have vigorously explored a variety of alternative emission mechanisms (see \citealt{mrm21a} for a recent critical review).
None of the alternatives have yielded a comparably compelling explanation for the linear polarization behavior
that is observed in many, but not all, pulsars (e.g. \citealt{mitra17}).

Our starting
point is the observation that important features of the radio emission -- its intrinsic stochasticity over a handful of rotations
and the detection of collective behavior within the pulse profile
\citep{gs03}  -- have never been incorporated convincingly into a
time-dependent model of the relativistic particle flow and electron-positron pair creation in the pulsar polar cap.   There is
a simple reason for this:  the dynamical time of charged particles in a surface gap
\citep{rs75,as79,mt92}  is several orders of magnitude shorter than the 
rotation period in all but the most slowly rotating pulsars.  Although the timescale of
  ${\bf E}\times{\bf B}$ drift in the polar cap can be longer than a rotation period
  \citep{gil03,basu20}, the phenomenon of sub-pulse drift also implies 
  a strong, long-lived angular inhomogeneity in the plasma discharge -- extending
  over $\sim 10^4$ repetitions (or more).   The case for such durable angular structure
  has remained nebulous.   No quantitative explanation has been given for how it would
    be created by the interaction of magnetospheric plasma with the neutron star surface; nor is it
  yet seen in time-dependent particle-in-cell (PIC) calculations \citep{levinson05,philippov20,cruz21}.

Another starting observation is that current carrying, magnetized plasmas are famously susceptible to internal resistive instabilities.
An elaborate theory of these instabilities has been developed to understand plasma (de-)confinement in compact fusion devices
\citep{white13}.
The profile of the current that initially flows through the plasma is not, in general, the profile that the plasma wishes to support.
In a tokamak, for example, a current profile initially peaked off the main toroidal axis will quickly relax to one peaked on axis.  
This relaxation appears to be mediated, in part, by high-order tearing modes, which feed off local peaks in the twist profile.

In a pulsar, a non-potential ($\sim$ azimuthal) magnetic field is established in the polar cap by the outward corotation charge flow
(at least, in neutron stars rotating rapidly enough to sustain electron-positron pair creation: \citealt{chen14,philippov14}).
This azimuthal field is
relatively weak, which implies that internal tearing instabilities are slow, small scale, and hard to resolve in
global, three-dimensional PIC simulations that cover the entire open circuit of a rotating
neutron star.

\subsection{Slow Tearing of a Quantizing Magnetic Field}

Existing kinetic treatments of magnetic tearing in relativistic
plasmas are not applicable to the pulsar polar cap, where the magnetic field is quantizing, charged particles are restricted to the lowest
Landau state, and curvature drift effects are completely negligible.   A kinetic theory of tearing in a quantizing magnetic field with 
relativistic particle flow is developed in \citep[hereafter Paper I]{t21}.   There we show that the tearing modes grow over the 
rotation period of the neutron star, and grow fastest below the magnetospheric skin depth.  What is more,
these modes are overstable, with a finite real frequency, in cases where the plasma carries net charge
(as it must in the pulsar magnetosphere: \citealt{gj69}).  In other words, the phenomenon
of magnetic tearing in the open pulsar circuit automatically incorporates the effect of azimuthal drift.
This result immediately suggests that 
the angular drift of radio sub-pulses is tied to coherent structures in the magnetic field, and not primarily
to rapid, evanescent oscillations in charged particles and electric field (the `sparks'
suggested originally by \citealt{rs75}).

In summary, our basic proposal is that the magnetic field in the open pulsar circuit is not a passive
actor in the process of radio emission, although the field is much less dynamic than is seen
near the equatorial current sheet in global simulations (e.g. \citealt{cerutti17}).
The corotation charge flow deposits energy in two reservoirs:  the kinetic energy of the charges and
the non-potential magnetic field.  The premise here is that repeated magnetic tearing induces a cascade
toward higher perpendicular wavenumber $k_\perp$ in both of these reservoirs.

\subsection{Cerenkov Emission of Localized Alfv\'en Modes}

In this picture, a secondary instability is required to trigger radio emission.
There is a long history of positing that longitudinal (Langmuir)
waves will be excited in the presence of a time-dependent momentum distribution of the charge
flow \citep{rs75,usov87,asseo90,mgp00}.  
But the nearly force-free state of the magnetic field
and the slow growth of the tearing mode together ensure that the current has a weak gradient
along the magnetic field.  Our focus is, instead, on an instability feeding off the transverse gradient
in the particle flow.

In this paper, we demonstrate a linear kinetic instability of Alfv\'en-like modes that are localized near a current sheet with a small thickness.  The underlying mechanism is simple.  In a uniform plasma, a shear Alfv\'en wave (a wave with a component $k_\perp$ of its wavevector perpendicular to the magnetic field) is slowed down significantly when $k_\perp \sim k_p
= \omega_p/c$, where $\omega_p$ is the plasma frequency.  No overstability is encountered if the charged particles supporting the current
flow uniformly along the background magnetic field: in such a situation one may Lorentz transform to the rest frame of the charges, and so the wave always propagates more rapidly than the particles.
But Cerenkov emission is possible when the charge flow varies across the magnetic field on a lengthscale
comparable to the skin depth $\sim k_p^{-1}$.
The particle beam in the current sheet outruns a shear Alfv\'en wave with $k_\perp$ comparable to the inverse
width $1/\Delta$ of the sheet.  The excited mode is localized near the position of the exciting beam,
and efficient growth requires a particle Lorentz factor $\bar\gamma_0 \lesssim 10^{2-3}$.
The mode is only approximately resonant with the beam, due to its spatial localization and strong growth, properties
  which are directly related.

Previous efforts to obtain a Cerenkov instability of shear Alfv\'en waves \citep{lominadze82,lyutikov00} have involved
introducing a high-momentum beam that moves super-Alfv\'enically with respect to a background plasma.
In this approach, the conversion of the generated subluminal Alfv\'en waves to superluminal radio waves has remained
an open question.
Our focus on lateral inhomogeneities in the current is motivated, first, by their
presence in the pulsar polar cap on scales comparable to the cap width \citep{timokhin13,gralla17}.  One
generically finds local extrema in the twist profile, which are promising sites for the
excitation of smaller-scale current gradients by the magnetic tearing process.  This allows a simple
longitudinal (one-dimensional) phase space distribution of the charged particles, in comparison with
the two-component beam model.  The lateral structure in the background
current also allows the Alfv\'en modes to be spatially localized.

\subsection{Related Three-dimensional Effects}

A recent proposal by \cite{mrm21b}, that
pulsar magnetospheres contain overdense `fibers', also relies on transverse structure,
but now involving variations in plasma density
rather than current.  Variations in current may be more natural on the plasma skin depth,  as considered here,
given that the conversion of gamma rays to secondary $e^\pm$ pairs typically involves the propagation
of a gamma ray over a much greater distance transverse to the magnetic field.

In addition, PIC simulations of time-dependent gaps, recently incorporating
pair creation \citep{philippov20,cruz21}, reveal the
excitation of transversely propagating electromagnetic waves
in the presence of weakly coordinated longitudinal oscillations.   The loss of coordination
on magnetic field lines
separated by more than a Debye length also leads to the stochastic excitation of subluminal Alfv\'en waves
(e.g. \citealt{bt07}).  
The recent PIC simulations provide a confirmation that a time-dependent gap structure will produce a quasi-steady
flux of radio waves above the plasma cutoff.  Nonetheless, this picture fails to account in a straightforward
manner for the stochasticity in the radio emission seen over several rotation periods and for detections of
collective sub-pulse drift.  In contrast with a longitudinal maser instability (e.g. \citealt{schopper03}),
there is no preferred orientation of the emitted electric vector with respect to rotations about
the background magnetic field. 

Finally, we note the old demonstration that shear in the particle momentum flux across the pulsar polar cap
can drive a transverse electromagnetic mode, the instability being
mediated by ${\bf E}\times{\bf B}$ drift \citep{asmith79}.  Since growth is suppressed by the
inverse of the very strong poloidal magnetic field, a transverse mode wavenumber much larger than $k_p$
is required, even if the particle energy approaches the maximum that can be supplied by the
voltage of the open pulsar circuit.  By contrast, the instability described here can operate both
in a surface gap and also within the pair creation zone above it, where the mean $e^\pm$ kinetic energy is
much lower.

\begin{figure}
  \epsscale{0.85}
  \vskip -0.5in
  \plotone{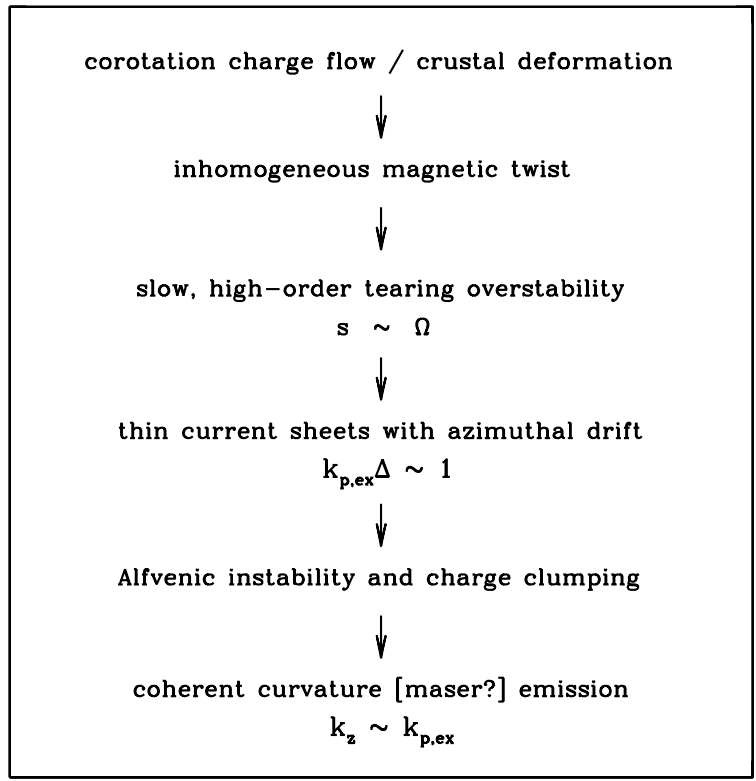}
  \vskip -1.3in
  \caption{Sequence of two physical instabilities that are proposed to trigger pulsar radio emission:
    1. Slow, high-order tearing in a quantizing
    magnetic field (Paper I).
    This induces relativistic charge flow in thin current sheets, $k_{p,\rm ex}\Delta \sim 1$,
    where $k_{p,\rm ex}^{-1}$ is the ambient skin depth.
    2. Alfv\'enic overstability of the current sheet, leading to charge clumping and coherent emission
    in the radio band by curvature radiation and (possibly) longitudinal maser emission.\label{fig:flowchart}}
\end{figure}

\subsection{From Slow Magnetic Tearing to Coherent Radio Emission}\label{s:flow}

Figure \ref{fig:flowchart} presents a flow chart starting with the corotation charge flow, leading
to the excitation of an inhomogeneous magnetic shear in the open pulsar circuit, and then to the
excitation of Alfv\'en waves trapped near thin current sheets, of thickness $\Delta$ comparable to
the ambient skin depth.

As is shown in Paper I,
a current distortion with wavenumber $k_x \sim k_p$ transverse to the guide magnetic field $B_{z0}\hat z$ experiences a
tearing instability with wavenumber $k_y$ and growth rate $\gtrsim 2\Omega (k_y/k_z)$.  Here $\Omega$
is the angular frequency of rotation of the star.  A single current sheet of
thickness $\Delta \lesssim k_p^{-1}$ experiences a somewhat faster tearing instability, with growth rate
diverging as $\sim \Delta^{-1/2}$.
The tendency for a nearly force-free magnetic configuration with a strong guide field
to form strong, localized current sheets is long familiar from the work of \cite{syrovatskii71}, and has been seen in
direct PIC simulations of tearing growth in a non-quantizing magnetic field with finite curvature drift
(e.g. \citealt{zenitani08,hoshino21}).

Each thin current sheet is unstable to a broad spectrum of trapped Alfv\'en waves.
This instability opens up two channels leading to the formation of macroscopic clumps of electric charge.
First, the Alfv\'en modes, which involve a growing charge perturbation, may directly combine into charged solitons.
These solitons would, like the linearly unstable waves, be a hybrid of a transverse (electromagnetic)
and a longitudinal (Langmuir-like) excitation.  The linear mode structure accounts self-consistently
for the effects of Debye screening, indicating that unscreened curvature emission is possible at low frequencies.

Another possibility is that a secondary charge density wave is excited in the primary particle
beam through its interaction with the trapped Alfv\'en mode.  The radiation by such a secondary charge
density wave can be viewed as a form of induced scattering \citep{lyubarskii96}.
Due to the hybrid nature of the Alfv\'en mode, this secondary charge density wave would be a hybrid
of the ``longitudinal maser'' discussed by \cite{schopper03} and the ``free electron maser''
discussed by \cite{fung04} and \cite{lyutikov21} that feeds off the transverse electric field.
The longitudinal electric field component has the dominant effect,
as there is no suppression of the induced plasma polarization by
the strong background magnetic field.  The transverse electric field only excites slow ${\bf E}\times{\bf B}$
drift of the $e^\pm$.  This drift has the same sign for $e^+$ and $e^-$, with the consequence that
a high pair multiplicity causes no amplification of the polarization compared with a pure corotation charge flow.
Finally, when charged solitons are present, the primary beam may interact with them
to produce a high-frequency tail to the emission spectrum -- an effect that is seen experimentally
in unmagnetized plasmas (e.g. \citealt{kato83}).

Extraordinary-mode (X-mode)
curvature radiation by Alfv\'en solitons is the default emission mechanism examined here;  see
Section \ref{s:curve} for details.
The longitudinal maser process is more tightly constrained by the plasma cutoff, especially when
a conservative account is made of the growth of seed linear waves. 
The relative challenges facing the maser mechanism are described in more detail in
Section \ref{s:maser}.

\subsection{Plan of the Paper}

The plan of this paper is as follows.  Section \ref{s:cerenkov} reviews relativistic shear Alfv\'en waves and
summarizes the results of our kinetic analysis of trapped Alfv\'en modes. Section \ref{s:kinetic} includes
a detailed derivation of the mode dispersion relation and group speed, as well as 
a qualitative discussion of mode saturation.
The radiative properties of nonlinear trapped Alfv\'en wavepackets are addressed in Section \ref{s:curve}.
Here, the focus is on the constraints imposed by rapid mode growth;  at emission, curvature radiation
is entirely polarized in the X mode, with a peak frequency far below the
ambient plasma frequency.  Broader
implications for pulsar and magnetar radio emission are drawn in Section \ref{s:pulsars},
including estimates of the implied pair multiplicity and the angular drift of the magnetic structures
created by tearing.  Two Appendices are devoted to deriving the peak growth rate of the Cerenkov instability
and describing the effect of bulk particle streaming outside the current sheet.

Throughout this paper, we adopt the shorthand $X = X_n\times 10^n$ to describe the normalization of quantity $X$
in c.g.s. units.

\section{Alfv\'enic Overstability of a Thin, Relativistic Current Sheet}\label{s:cerenkov}

We start with the observation that a shear Alfv\'en wave is slowed dramatically
when its wavefront varies significantly over a distance comparable to the plasma skin depth in
a direction perpendicular to the background magnetic field (${\bf B}_0 = B_{z0}\hat z$). The
simplest case is a cold and homogeneous plasma with an extreme magnetization $\sigma = B^2_{z0}/4\pi n_0 mc^2 \gg 1$;
the wave dispersion relation reads (\citealt{ab86,bellan06}; Section \ref{s:alfven})
\be\label{eq:alfven}
\omega({\bf k}) = {ck_\parallel\over\sqrt{1 + k_\perp^2/k_p^2}}\quad\quad(\omega \ll ck_p).
\ee
Here, $k_\parallel = {\bf k}\cdot \hat z$ is the parallel wavenumber and
$k_p \equiv \omega_p/c = (4\pi n_0 q^2/mc^2)^{1/2}$ the inverse skin depth {\bf associated with
mobile charges of number density $n_0$ and charge $q$}.
More generally, the plasma particles may have a finite dispersion or bulk motion along ${\bf B}_0$,
which is taken to be strong enough to induce rapid radiative transitions to the ground Landau state.

The reduction in the phase speed $\omega/k_\parallel$ leads us to consider whether a Cerenkov-like instability
is possible in an inhomogeneous plasma.  If the bulk speed of the plasma particles varies in a
direction perpendicular to ${\bf B}_0$, it may in some places exceed $\omega/k_\parallel$.
The particular structure investigated here is a thin
current sheet, of half-thickness $\Delta \sim k_p^{-1}$.  Such structures are naturally produced
in a quantizing magnetic field via the slow, resistive instability investigated in Paper I.

The shear Alfv\'en wave has a transverse (electromagnetic) component, as well as a longitudinal
(electrostatic) component.  The two components of the wave have comparable energy densities
when $k_\perp \sim k_p$, but only the transverse component transports energy through
a static plasma.  Such a hybrid wave couples readily to a sheared background
particle flow, even in the case of extreme magnetization where transverse ${\bf E}\times{\bf B}$
particle drift is suppressed.

In what follows, the mobile
charges are predominantly electron-positron pairs in the parts of a pulsar magnetosphere sustaining
pair creation, but could either be ions or electrons in the parts where the plasma is sourced
only by the corotation charge flow.  The instability uncovered here may operate in both regimes
of a pair-rich and a pair-starved current.

\subsection{Homogeneous Plasma}\label{s:homog}

It is worth reviewing why
the Cerenkov excitation of shear Alfv\'en waves by bulk plasma motion is not possible in a
homogeneous medium.  One can always boost to the plasma rest frame, where
the dispersion relation (\ref{eq:alfven}) is recovered, meaning that the wave phase speed always
exceeds the bulk speed $\bar\beta_0 c$.  For example, Lorentz transforming the dispersion relation
(\ref{eq:alfven}) by a velocity $-\bar\beta_0c$ gives the modified Alfv\'en speed
\be
\beta_A c \;=\;  {c\over\sqrt{1 + k_\perp^2/k_p^2}} \quad\rightarrow \quad \beta_A'c - \bar\beta_0c
\;=\; {\beta_A c\over \bar\gamma_0^2(1 + \beta_A\bar\beta_0)} > 0.
\ee
Here, $k_\perp$ is invariant under the boost and $k_p$ is defined in the original static plasma frame.  The
equivalent phase Lorentz factors $\gamma = (1-\beta^2)^{-1/2}$ are related by
\be
\gamma_A' = \bar\gamma_0(1+\beta_A\bar\beta_0)\gamma_A > \bar\gamma_0.
\ee

There is a contrast here with the 
interstellar medium, where a heavy, cold, background plasma is present
and the streaming speed of cosmic rays may be limited to the non-relativistic Alfv\'en speed by
the Cerenkov excitation of Alfv\'en waves \citep{kulsrud69}.  This streaming instability re-emerges in the
pulsar magnetosphere if one introduces a second longitudinal (beam) momentum component to the plasma
\citep{lominadze82,lyutikov00}; however, the origin of the beam, the transverse structure of
the excited shear Alfv\'en wave, and its coupling to an escaping superluminal mode have never
been clearly defined.

\vfil\eject
\subsection{Alfv\'en-like Modes Trapped Near Thin Current Sheets:  Summary of Results}

We consider Alfv\'en-like modes that are localized near a current sheet of a small thickness
$\Delta$, comparable to the ambient skin depth $k_{p,\rm ex}^{-1}$,
embedded in a quantizing magnetic field.  These modes propagate below the plasma frequency,
carry charge and current perturbations, and have a group velocity aligned with the local guide magnetic field.
These trapped Alfv\'en modes are supported by scalar and vector potential perturbations $\{\phi_1, A_{z1}\}$,
with eigenfunction $\propto f(x)e^{st + i(k_zz-\omega t)}$ and $f(x)$ decaying exponentially
away from the sheet, at $|x| > \Delta$.

Charges within the sheet flow relativistically along the magnetic field
with Lorentz factor $\bar\gamma_0$.
Because the particle flow speed differs inside and outside the current sheet,
  the phase speed of the excited mode is resonant with neither particle component,
  adjusting to an intermediate value.  The default background configuration has vanishing
current outside the sheet (effectively, a large ratio of current density within the sheet
relative to the background corotation current in its exterior).
A straightforward generalization of the model allows for relativistic bulk
motion outside the sheet, associated with a secondary pair cascade,
with Lorentz factor $\bar\gamma_{0,\rm ex} \lesssim \bar\gamma_0$.

The dispersion relation of trapped Alfv\'en modes
is summarized in Section \ref{s:rel}, in particular in Equations (\ref{eq:mode0}),
(\ref{eq:mode1}) and (\ref{eq:mode2}) and Figure \ref{fig:dispersion}, \ref{fig:velocity}.
The modes have the following properties.

\vskip .1in
\noindent
1. The mode slightly lags the bulk particle flow.  Its phase and groups speeds are fully relativistic
when $\bar\gamma_0 \gg 1$, in contrast with a shear Alfv\'en mode with $k_\perp \sim k_p$.
The lag is $\bar\beta_0c - \omega/k_z \sim c/\bar\gamma_0^2$, and the growth rate
is $s \sim ck_{p,\rm ex}/\bar\gamma_0^2$.
\vskip .1in
\noindent
2. The growth rate and the phase speed lag depend on the parameter
$\varepsilon \equiv (k_{p,\rm ex}\Delta)^2/4(k_z^2/k_{p,\rm ex}^2-1)$.  Growing modes are found only for $k_z < k_{p,\rm ex}$,
meaning that the modes have a longitudinal wavelength comparable to the skin depth outside the current sheet.
The penetration depth outside the current sheet is larger by a factor $\sim \bar\gamma_0$
(Equation (\ref{eq:realkap})).
\vskip .1in
\noindent
3. Although growth rates rise for $|\varepsilon| \gg 1$, the group speed also drops
(Equations (\ref{eq:vgr}), (\ref{eq:gamgr}) and Figure \ref{fig:velocity}).  The modes
that are the most promising seeds for charge clumping and curvature radiation are those with $|\varepsilon| = O(1)$.
\vskip .1in
\noindent
4. The localization of the Alfv\'en-like mode near the current sheet is closely tied to its overstability
(Equation (\ref{eq:realkap})).  The Poynting flux carried by the mode is directed along
  the strong guide magnetic field.
By contrast, the Poynting flux carried by a plane- or cylindrically-polarized shear Alfv\'en wave
diverges over the plane transverse to ${\bf B}_0$ (Section \ref{s:homog}).    A plane-polarized
electromagnetic wave propagating through a homogeneous magnetized plasma may have an arbitrary profile
in the transverse plane only in the force-free approximation \citep{tb98,gralla14}, which ignores
the longitudinal dynamics that is central to the Cerenkov instability.
\vskip .1in
\noindent
5. Rapid mode growth is possible, with the mode structure adjusting to long-range gradients in plasma density
(Section \ref{s:curve}).
This requires a mean particle Lorentz factor $\bar\gamma_0 \sim 30-100$, comparable
to that expected from a pair cascade in the polar cap of a radio pulsar (Figure \ref{fig:omcpk}).
Then the mode group Lorentz factor $\gamma_{\rm gr}$
is small enough that a nonlinear wavepacket behaves like a point charge
as it moves along the curved polar magnetic field of a neutron star (Section \ref{s:size}).
\vskip .1in
\noindent
6. The kinetic equations show that mode growth is driven by the bulk particle flow, not by the current.
The mode energy is dominated by the transverse electromagnetic field when $|\varepsilon| = O(1)$
(Section \ref{s:sat}).  The energy carried by the longitudinal component
is smaller by a factor $\sim 1/\bar\gamma_0$.  Mode growth is argued to saturate when the transverse field energy
becomes comparable to the kinetic energy of the background particle flow within the current sheet.
\vskip .1in
\noindent
7. A further consequence of the low $\gamma_{\rm gr}$ needed for rapid growth
is that the peak curvature frequency generated by non-linear wavepackets is limited to about $10^{-2}$ of
the ambient plasma frequency $\omega_{p,\rm ex} = ck_{p,\rm ex}$ (Figure \ref{fig:omcpk}).
Curvature radiation is then polarized entirely in the X mode.  The net radiated power
over the radiation decoupling length $R_c/\gamma_{\rm gr}$ (here $R_c$ is the magnetic radius of curvature)
is limited to $\sim 10^{-4}$ of the kinetic energy of the charge flow through the current sheet
(Figure \ref{fig:power}).  This rises to $\sim 10^{-3}-10^{-2}$ over the radius $r$.
\vskip .1in
\noindent
8. The charge carried by the mode is dominated by the medium outside the current sheet (Section \ref{s:charge}
and Figure \ref{fig:charge}).  This is a key 
feature as regards the radiation of extraordinary waves below the ambient plasma frequency: the charge
profile is established self-consistently including the effect of Debye screening by the ambient medium.

\vfil\eject
\section{Kinetic Derivation of the Dispersion Relation}\label{s:kinetic}

To set the stage for our analysis of Cerenkov instability, we first re-derive the dispersion relation
of a shear Alfv\'en wave propagating through a homogeneous, static plasma in a quantizing magnetic field.  This is followed by a kinetic
treatment of Alfv\'en-like excitations localized around a thin current sheet.  The eigenvalue equation
is solved analytically in two cases:  slow, subrelativistic particle drift in the sheet, and relativistic
bulk flow.  The second case, which is relevant to radio curvature emission from a pulsar, is analyzed in some detail.  

\subsection{Shear Alfv\'en Wave in a Homogeneous Plasma:  Review}\label{s:alfven}

The plasma is embedded in a uniform magnetic field ${\bf B}_0 = B_{z0}\hat z$ and
its magnetization $\sigma$ is extreme.  In the context of pulsars, a particle of electric charge $q$ and mass $m$
has a cyclotron frequency $qB_{z0}/mc = \sigma^{1/2} \omega_p \sim 10^{8-10}\omega_p$.
The charge experiences negligible
transverse ${\bf E}\times{\bf B}$ drift in response to a propagating electromagnetic disturbance, the drift rate being
suppressed by a factor $\sim \sigma^{-1}$.  The particle dynamics is then entirely longitudinal.

An electric four-current $(\rho_1, J_{z1})$ is excited when the wave varies in the coordinates
${\bf x}_\perp = (x,y)$ perpendicular to ${\bf B}_0$, implying that the electromagnetic field has both
transverse and longitudinal components.  The Alfv\'en wave is most compactly described by the vector
potential $A_{z1}({\bf x}_\perp,z,t)$ and electrostatic potential $\phi_1({\bf x}_\perp,z,t)$.
In a planar geometry (we work in Lorentz gauge),
\be\label{eq:planar}
A_{z1}({\bf x},t) = \widetilde A_{z1}\,e^{i({\bf k}\cdot{\bf x} - \omega t)};\quad  \phi_1 = {ck_z\over\omega}A_{z1}.
\ee
The wave vector ${\bf k} = ({\bf k}_\perp, k_z)$ has arbitrary orientation, and the field components are
\be
   {\bf B}_{\perp1} = iA_{z1}\,{\bf k}_\perp\times\hat z; \quad  {\bf E}_{\perp1} = -i{\bf k}_\perp\phi_1;
\ee
\be
   E_{z1} = i\left({\omega\over c}A_{z1} - k_z\phi_1\right) = i{\omega^2 - c^2k_z^2\over \omega c}A_{z1}.
\ee
For reference, the cylindrically symmetric waveform, which satisfies the same dispersion relation, is
\be\label{eq:cyl}
A_{z1}(R,z,t) = \widetilde A_{z1}\, J_0(k_\perp R) e^{i(k_z z - \omega t)}.
\ee
The transverse field components are replaced by
\be
B_{\phi1} = -k_\perp \widetilde A_{z1} J_1(k_\perp R) e^{i(k_z z - \omega t)};  \quad
E_{R1} = -k_\perp \widetilde \phi_{z1} J_1(k_\perp R) e^{i(k_z z - \omega t)}.
\ee

The dispersion relation is obtained from the wave equation for $E_{z1}$,
\be\label{eq:wave}
\left({1\over c^2}{\partial^2\over\partial t^2} - {\partial^2\over\partial z^2} - \nabla_\perp^2\right)E_{z1}
=
4\pi i\left({\omega\over c^2}J_{z1} - k_z\rho_1\right) = -k_p^2\left(1 - {c^2k_z^2\over\omega^2}\right)E_{z1}.
\ee
Here, $\nabla_\perp^2 \sim -k_\perp^2$ is the transverse Laplacian.
The current perturbation is obtained from the longitudinal force equation, and the charge density perturbation
from considerations of charge conservation,
\be\label{eq:cur1}
J_{z1} = i {n_0q^2 \over m\omega}E_{z1} = i{k_p^2c^2\over 4\pi \omega}E_{z1};  \quad  \rho_1 = {k_zc\over\omega}J_{z1}.
\ee
Equation (\ref{eq:wave}) implies
\be
(k_p^2c^2 - \omega^2)(c^2k_z^2 - \omega^2) = \omega^2 c^2k_\perp^2.
\ee
The Alfv\'en mode is the lowest-frequency solution to this equation, simplifying to
Equation (\ref{eq:alfven}) when $k_z \ll k_p$ but $k_\perp$ has arbitrary magnitude.

It should be emphasized that this derivation of the shear Alfv\'en mode
  does not depend on the balance between positive and negative charges in the plasma: Equation
  (\ref{eq:cur1}) is independent of the sign of $q$.  There is an implicit assumption that
  the background parallel electric field $E_{z0}$ is small enough to have a negligible effect
  on the particle dynamics over a wave period.  Thus, high-frequency Alfv\'en waves can be supported
  in parts of the pulsar polar cap where $E_{z0}$ is limited by screening -- either because
 $|J_{z0}| < |\rho_{\rm co}| c$ \citep{timokhin13}, or because $e^\pm$ pairs are created with high multiplicity.

\subsection{Overstability of Trapped Alfv\'en Modes at Thin Current Sheets}\label{s:overstable}

We now generalize the eigenvalue problem in two ways.  First, a weak background current is introduced, localized to
a thin current sheet extending over $-\Delta < x < \Delta$.  The strong `guide' magnetic field
$B_{z0}(x) \simeq {\cal B}_{\parallel0}$ is weakly sheared in a transverse direction $\hat y$,
\be\label{eq:B0}
         {\bf B}_0(x) = B_{z0}(x)\hat z + B_{y0}(x)\hat y \simeq
         \left[{\cal B}_{\parallel0} - {B_{y0}^2(x)\over 2{\cal B}_{\parallel0}}\right]\hat z + B_{y0}(x)\hat y
         \quad\quad({\cal B}_{\parallel0} = {\rm const}).
\ee
The current is assumed to vanish outside the sheet,
\be\label{eq:Jz0}
J_{z0} = {cB_{\perp0}\over 4\pi\Delta} \quad (-\Delta < x < \Delta);  \quad\quad  J_{z0} = 0 \quad (|x| > \Delta).
\ee

Second, relativistic bulk plasma motion is introduced within the current sheet.
The charges flow along the magnetic field with a mean Lorentz factor $\bar\gamma_0$.  They impart a
minuscule stress, as small as $\sim 10^{-17} {\cal B}_{\parallel0}^2/4\pi$ for a typical radio pulsar.
The magnetic field can therefore be taken to be force-free.
In slab geometry, this corresponds to $(d/dx)(B_{y0}^2 + B_{z0}^2) = 0$, whence the expansion of $B_{z0}$ in
Equation (\ref{eq:B0}).

The negative and positive charges are taken to have the same phase space distribution but with different
overall densities.  We choose a distribution function with a narrow spread in momentum space;
  a concrete example is a top hat centered at $\bar p_0 \equiv \bar\gamma_0\bar\beta_0 mc$,
\be\label{eq:f0}
f_0^\pm(p) = {n_0^\pm\over \Delta p_0}\Theta(p-p_{0-})\Theta(p_{0+}-p).
\ee
Here, $p_{0\pm} = \bar p_0 \pm \Delta p_0/2$ and $\Theta$ is the Heaviside function.\footnote{$\Theta(x) = 0$ $(1)$ for $x<0$ ($x>0$).}
As in Paper I, we assume that $\bar\gamma_0 \ll B_{z0}/B_{\perp0}$, implying $1-\bar\beta_0 \gg (B_{y0}/B_{z0})^2$.

The presence of a current implies an imbalance in the densities of positive and negative charges, $n_0^\pm = (1\pm\varepsilon_\rho)n_0/2$.
The total particle density,
\be
n_0(x) = n_0^+(x) + n_0^-(x) = {J_{z0}\over \varepsilon_\rho q\bar\beta_0 c},
\ee
is taken to be constant everywhere inside and outside the current sheet, $n_0(x) =$ constant.  The particle velocity therefore
vanishes at $|x| > \Delta$.

The simplest representation of this model in the pulsar magnetosphere is a current sheet with $J_{z0}$ approaching $\rho_{\rm co}c$
and $\varepsilon_\rho \simeq 1$ (no pairs inside or outside the sheet),
surrounded by a more extended zone with $J_{z0} \lesssim \rho_{\rm co}c/2$.
The charges in the sheet are forced to flow relativistically but, outside the sheet, the current can be supplied by a subrelativistic
drift of the corotation charge.  The charge density, measured by the parameter $\varepsilon_\rho$, is also taken to be constant.

When secondary pairs are present, the particle distribution function separates at first into a high-momentum
  beam composed of the primary corotation charge flow, and a secondary $e^\pm$ component.
  As we show below, the primary beam (Lorentz factor $\gamma \sim 10^{6-7}$)
  has too high a momentum to excite a trapped Alfv\'en mode with a significant growth rate.
  The secondary pairs ($\gamma \sim 30-100$) have approximately equal average momenta, the relative
  offset being small at high pair multiplicity.

  In the picture described here and in Paper I, the strongest current and highest particle
  energy are concentrated in narrow zones, of thickness $\Delta$ comparable to the ambient
  skin depth.  Screening of strong $E_\parallel$ inside such a current sheet by the conversion of curvature gamma
  rays into $e^\pm$ pairs requires a higher photon energy, as compared with a situation where the current
  is smoothly distributed across the pulsar polar cap.  The small value of $\Delta$ implies a reduced
  propagation distance $\ell_\pm$ between gamma-ray emission and pair conversion.
  A gamma ray propagating across magnetic field lines of curvature radius
  $R_c$ will leave the current sheet over a distance $\ell_\pm \sim (R_c \Delta)^{1/2} \sim 0.1\,(R_c/100~{\rm km})^{1/2}
  (\Delta/10~{\rm cm})^{1/2}$ km.  Lower-energy curvature gamma rays convert over a larger distance,
  and produce lower-energy pairs outside the sheet.  Hence, the average momentum of the secondary pairs
  can be expected to peak within a current sheet.
  
  In this situation, relativistic bulk streaming is also present outside the current sheet.
  Due to the presence of an intense guide magnetic field, this situation may be represented by 
boosting to the rest frame of the medium outside the current sheet (Section \ref{s:streamext}, Appendix
\ref{s:streamext2}).  Although the weak transverse magnetic field near
the current sheet is not invariant under this boost, it does not enter into the kinetic equations derived below.
That is because the mode wavevector ${\bf k}$ has a significant component parallel to the guide field, so that
${\bf k}\cdot{\bf B}_0 \simeq k_z B_{z0}$.

The Alfv\'enic overstability and the more familiar two-stream instability tap independent sources of
  energy, connected with the differential in particle momentum across magnetic field lines, versus dispersion
  in momentum along a given magnetic flux element.  The relative magnitude of these energy sources must be 
  gauged using a detailed
  model of $e^\pm$ pair creation in magnetospheric current sheets, and will be addressed elsewhere.


\subsection{Perturbation Equations}

In contrast with Paper I, our analysis focuses on electromagnetic perturbations with finite longitudinal wavevector $k_z$.
A Cerenkov instability of shear Alfv\'en-like waves is uncovered in the simplest case $k_y = 0$, and so we make that restriction.
The perturbation to the electromagnetic field contains both electrostatic and longitudinal vector components ($\phi_1$ and
$A_{z1}$), which depend on $x$, $z$, and $t$,
\ba
   {\bf B}_\perp &=& {\bf B}_{\perp0} + \bnabla A_{z1}(x)\times \hat z =
   \left(B_{y0} - {\partial A_{z1}\over\partial x}\right)\hat y; \nn
   {\bf E}_{\perp1} &=& -{\partial\phi_1\over\partial x}\hat x;\quad\quad
   E_{\parallel,1} \equiv {\bf E}_1\cdot\hat B \simeq -{1\over c}{\partial A_{z1}\over\partial t} - {\partial\phi_1\over\partial z}.
\ea
In the background state, the particle distribution function depends on $x$ and the kinetic momentum $p \simeq p_z$
parallel to the magnetic field.  One has
\be
f(x,z,p,t) = f_0(x,p) + f_1(x,z,p,t)
\ee
and the perturbation particle velocity is (Paper I)
\be\label{eq:betaperp}
\bbeta = \bbeta_0 + \bbeta_1 \simeq \beta_0\left(\hat z + {{\bf B}_{\perp 0}\over B_{z0}}\right) +
       {{\bf E}_{\perp1}\times {\bf B}_{z0}\over B_{z0}^2} + \beta_0{{\bf B}_{\perp 1}\over B_{z0}} +
       \beta_1{{\bf B}_{\perp 0}\over B_{z0}}.
       \ee
Each fourier component decomposes as
\be\label{eq:fourier}
A_{z1} = \widetilde A_{z1}(x) e^{(s-i\omega)t + ik_zz}; \quad\quad f_1 = \widetilde f_1(x) e^{(s-i\omega)t + ik_zz}.
\ee
Our goal is to calculate the mode growth rate $s$ and real frequency $\omega$ as functions of $k_p\Delta$.
In the process, we will also obtain the transverse eigenfunctions $\widetilde A_{z1}(x)$, $\widetilde f_1$.

The perturbed Boltzmann equation reads
\be\label{eq:Boltzperp}
   {\partial f_1^\pm\over\partial t} + c\bbeta_0^\pm\cdot\bnabla f_1^\pm + c\bbeta_1^\pm\cdot\bnabla f_0^\pm =
   \mp q({\bf E}_1\cdot\hat B){\partial f_0^\pm\over \partial p},
\ee
where $\pm$ labels positive and negative charges.  Because (i) the wavevector is directed along the guide field,
and (ii) ${\bf B}_{\perp 0}$ has no component in the gradient direction $\hat x$ of the background, we have
${\bbeta}_0^\pm\cdot{\bf k} = \beta_0 k_z$.
Substituting Equation (\ref{eq:betaperp}) into Equation (\ref{eq:Boltzperp}) gives
\be\label{eq:f1}
f_1^\pm = \mp {q\over s-i(\omega - \beta_0ck_z)}E_{\parallel,1}{\partial f_0^\pm\over\partial p}
\ee
and the longitudinal electric field is now
\be\label{eq:Epar}
E_{\parallel,1} = -\left({s-i\omega\over c}A_{z1} + ik_z\phi_1\right)
= -\left({s-i\omega\over c} + {ck_z^2\over s-i\omega}\right)A_{z1}.
\ee
The potentials are related by $\phi_1 = -ick_z(s-i\omega)^{-1}A_{z1}$.  
Because $k_y = 0$, the longitudinal current satisfies the conservation law
\be\label{eq:cons}
(s-i\omega)\rho_1 + ik_z J_{z1} = 0,
\ee
as may be seen by integrating the quantity $q(f_1^+ - f_1^-)$ over $p$.

  We search for modes whose phase speed $\omega/k_z$ along the strong, guide
  magnetic field is offset from the mean particle speed $\bar\beta_0 c$, the offset being larger
  than the velocity spread of the beam.
The perturbed current density is then readily obtained from Equations (\ref{eq:f0}) and (\ref{eq:f1}),
\be\label{eq:Jz}
   4\pi J_{z1} = 4\pi q\int dp \beta_0(p) c[f_1^+(p) - f_1^-(p)] = 
     k_p^2c^2 {s-i\omega \over (s-i\bar\omega)^2} E_{\parallel,1},
\ee
where
\be
\bar\omega \equiv \omega - \bar\beta_0 ck_z
\ee
is the Doppler-shifted frequency evaluated at the velocity center of the
  beam.\footnote{This result does not depend on the detailed shape of the distribution
  function -- e.g. top hat vs. Gaussian -- as long as it is narrow in momentum space.}
The charge density perturbation $\rho_1$ follows from Equation (\ref{eq:cons}).  The skin depth inside the current sheet,
as determined by the integral (\ref{eq:Jz}), is
\be
k_{p,\rm in}^2 = {4\pi n_0 q^2\over \bar\gamma_0^3mc^2} = {1\over\varepsilon_\rho\bar\beta_0\bar\gamma_0^3} {qB_{\perp0}\over mc^2\Delta},
\quad\quad(|x| < \Delta),
\ee
with
\be\label{eq:krel}
k_{p,\rm ex}^2 = 
\bar\gamma_0^3k_{p,\rm in}^2\quad\quad(|x| > \Delta)
\ee
outside the sheet.

One observes that, in contrast with the tearing mode studied in Paper I, the transverse non-potential
magnetic field has dropped out of the phase-space density perturbation $f_1^\pm$:  it influences the instability only indirectly through
the particle flow that sustains it.  The immediate energy source for the instability uncovered here
is therefore the kinetic energy of the charges.

The mode dispersion relation is obtained by combining the Maxwell equations for $\phi_1$ and $A_{z1}$
to give a wave equation for $E_{\parallel,1}$.   Starting from
\be       
\left[ {(s-i\omega)^2\over c^2} + k_z^2 - {\partial^2\over\partial x^2}\right]A_{z1} = {4\pi\over c}J_{z1},
\ee
and substituting Equation (\ref{eq:Epar}) on the left-hand side and (\ref{eq:Jz}) on the right-hand side gives
\be
\left[ {(s-i\omega)^2\over c^2} + k_z^2 - {\partial^2\over\partial x^2}\right]E_{\parallel,1}
  = -{4\pi\over c}\left({s-i\omega\over c} + {ck_z^2\over s-i\omega}\right)J_{z1}
  = -k_p^2{(s-i\omega)^2 + c^2k_z^2 \over (s-i\bar\omega)^2}\,E_{\parallel,1}.
  \ee
The current density is uniform both inside and outside the current sheet, and so the eigenvalue
equation of a single fourier mode (\ref{eq:fourier}) takes the simple form
\be\label{eq:eigen}
   {d^2\widetilde E_{\parallel,1}\over dx^2} = \kappa^2\widetilde E_{\parallel,1}.
   \ee
The coefficient $\kappa^2$ differs inside and outside the sheet.

\subsection{Formulation of the Eigenvalue Problem}

The solution to the eigenvalue Equation (\ref{eq:eigen}) may be either exponentially growing or decaying outside
  the current sheet.   We now find the most general smooth, symmetric solution with a finite energy, i.e., one that 
is localized around the current sheet and has a continuous derivative at $|x| = \Delta$.  This takes the form
\be\label{eq:Eparvsx}
\widetilde E_{\parallel,1} =
\begin{cases}
  \widetilde E_0\left(e^{\kappa_{\rm in}x} + e^{-\kappa_{\rm in}x}\right) & (|x| < \Delta);\\
  \widetilde E_0 \left(e^{\kappa_{\rm in}\Delta} + e^{-\kappa_{\rm in}\Delta}\right) e^{-\kappa_{\rm ex}(|x|-\Delta)} & (|x| > \Delta),
\end{cases}
\ee
where
\ba\label{eq:kap}
\kappa_{\rm in}^2 &=& \left[{(s-i\omega)^2\over c^2} + k_z^2\right]\left[1 + {c^2k_{p,\rm in}^2\over (s -i\bar\omega)^2}\right]
  \quad\quad (|x| < \Delta);\nn
\kappa_{\rm ex}^2 &=& \left[{(s-i\omega)^2\over c^2} + k_z^2\right]\left[1 + {c^2k_{p,\rm ex}^2\over (s -i\omega)^2}\right]
  \quad\quad (|x| > \Delta).
\ea
We define $\kappa_{\rm ex}$ as the root with positive real part.
The continuity of $d\widetilde E_{\parallel,1}/dx$ at $|x| = \Delta$ gives the eigenvalue equation
\be
\kappa_{\rm ex}  = -\kappa_{\rm in}{e^{\kappa_{\rm in}\Delta} -
  e^{-\kappa_{\rm in}\Delta} \over e^{\kappa_{\rm in}\Delta}+e^{-\kappa_{\rm in}\Delta}}.
\ee
When the current sheet is thin compared with the internal skin depth, this reduces to
\be\label{eq:disp}
\kappa_{\rm ex} \simeq - \kappa_{\rm in}^2 \Delta\quad\quad (|\kappa_{\rm in}|\Delta \ll 1).
\ee
We focus on solutions to this equation with ${\rm Re}[\kappa_{\rm in}^2] < 0$, corresponding to a decay
of the mode outside the sheet.  The thin-sheet approximation
is justified for sufficiently high $\bar\gamma_0$, as described in Section \ref{s:rel}.

\subsection{Mode Spectrum for Subrelativistic Particle Flow}\label{s:subrel}

The case of subrelativistic particle flow ($\bar\beta_0 \ll 1$) in a narrow current sheet ($k_{p,\rm ex}\Delta \ll 1$)
is simple and illustrative, even if the modes obtained are not directly relevant to pulsar radio emission.  We now
search for subluminal solutions, $\omega/k_z < c$, with frequencies below the ambient plasma cut-off,
$\omega \ll \omega_{p,\rm ex} = ck_{p,\rm ex}$.   It first should be noted that modes localized around the current sheet must
have non-vanishing growth rate, $s > 0$.  That is because the quantity $\kappa_{\rm ex}$ in Equation (\ref{eq:Eparvsx})
is purely imaginary when $s=0$.

We next recall that a shear Alfv\'en wave propagating with $k_\perp \gg k_{p,\rm ex}$ in a uniform medium is strongly
subluminal, $\omega/k_z \sim (k_{p,\rm ex}/k_\perp)c \ll c$.  This suggests that, in searching for
localized subluminal modes, we focus on the case $k_{p,\rm ex}\Delta \ll 1$.  The growing mode with
exponentially decaying amplitude outside the current sheet (${\rm Re}[\kappa_{\rm ex}] > 0$,
as defined in Equation (\ref{eq:Eparvsx})) has
\be\label{eq:kaplowbeta}
\kappa_{\rm ex} \;\simeq\; i\,{ck_{p,\rm ex}k_z\over \omega + is}; \quad\quad
  \kappa_{\rm in}^2 \;\simeq\; -{c^2k_{p,\rm ex}^2k_z^2\over (\omega - \bar\beta_0ck_z + is)^2}.
\ee
Normalizing
\be\label{eq:dimensionless0}
s \rightarrow \tilde s\cdot ck_z; \quad\quad \bar\omega = \omega - ck_z \rightarrow \varpi\cdot ck_z,
\ee
and substituting Equation (\ref{eq:kaplowbeta}) into Equation (\ref{eq:disp}) gives
\be\label{eq:displow}
\tilde s = - {\varpi + \bar\beta_0\over 2\varpi}k_{p,\rm ex}\Delta;\quad\quad
\varpi^4 = {\bar\beta_0^2-\varpi^2\over 4}(k_{p,\rm ex}\Delta)^2.
\ee

We deduce that the growing mode has a phase speed that lags the bulk speed of the particles,
$\omega/k_zc < \bar\beta_0$.  
Equations (\ref{eq:displow}) are easily solved in two regimes.  Strongest growth is found when $\bar\beta_0 \gg k_{p,\rm ex}\Delta$,
corresponding to a flow speed greater than the phase speed of a shear Alfv\'en wave with $k_\perp \sim 1/\Delta$.
Restoring dimensionful units,
\be
s \;\simeq\; \left({\bar\beta_0\over 2k_{p,\rm ex}\Delta}\right)^{1/2}\cdot (k_{p,\rm ex}\Delta) ck_z
\quad\quad (\bar\beta_0 \gg k_{p,\rm ex}\Delta),
\ee   
and the lag between phase speed and particle speed is small in amplitude,
\be
\bar\beta_0c - {\omega\over k_z}  \;\simeq\; \left({k_{p,\rm ex}\Delta\over 2\beta_0}\right)^{1/2}\,\bar\beta_0c \;\ll\; \bar\beta_0c.
\ee
A key feature of this solution is that the propagation of the mode has been `pulled up' to match closely
the particle flow.

As expected, growth is weaker in the opposing regime $k_{p,\rm ex}\Delta \gg \bar\beta_0$, and the phase speed is very small,
\be
s \;\simeq\; {\bar\beta_0^2\over k_{p,\rm ex}\Delta}ck_z; \quad\quad
{\omega\over k_z}  \;\simeq\; 2\left({\bar\beta_0\over k_{p,\rm ex}\Delta}\right)^{2}\,\bar\beta_0c
\quad\quad (\bar\beta_0 \ll k_{p,\rm ex}\Delta).
\ee
In both of these cases, the mode carries a current and charge perturbation, but its
phase and group speeds are subluminal and the radiative emissions are insignificant.

\subsection{Mode Spectrum for Relativistic Flow}\label{s:rel}

We next find a limiting solution to the eigenvalue problem in which the Lorentz factor $\bar\gamma_0$ of the
plasma in the current sheet is taken to be very high.  
The growing mode has a relativistic phase speed that, once again, turns out to lag slightly the particle speed;
both $s$ and $\bar\omega$ are of order $ck_z/\bar\gamma_0^2$.  
A true pole in the plasma response is absent simply because
$\omega/k_z$ is intermediate between the particle speeds inside and outside the supporting current sheet.
Mathematically, the rapid growth and the offset in phase speed are closely related, and both are
a consequence of the inhomogeneous mode structure in the $x$-direction.

Now $ck_{p,\rm in} \gg |s -i\bar\omega|$
and $\bar\gamma_0$ scales out of the eigenvalue equation, which will allow us to consider longitudinal
wavenumbers $k_z$ close to $k_{p,\rm ex}$.  The current density in the sheet simplifies to
\be\label{eq:Jzin}
   {4\pi\over c}J_{z1} \simeq -\kappa_{\rm in}^2 A_{z1}.
\ee
Setting aside for the moment the constraint
arising from the sign of $\kappa_{\rm in}^2$, we take the square of Equation (\ref{eq:disp}) to get
\be\label{eq:disp2}
{(\bar\omega + is)^4\over c^4}\left(1 - {k_{p,\rm ex}^2\over k_z^2}\right) =
  \left({k_z^2\over\bar\gamma_0^2} - 2k_z{\bar\omega + is\over c}\right){k_{p,\rm ex}^4\Delta^2\over\bar\gamma_0^6}.
\ee
Here, we have substituted the relations $\omega = \bar\omega + \bar\beta_0ck_z$ and
$k_{p,\rm in}^2 = k_{p,\rm ex}^2/\bar\gamma_0^3$ (from Equation (\ref{eq:krel})) and implemented $\bar\gamma_0 \gg 1$.
Equation (\ref{eq:disp2}) includes
the relevant terms to leading order in $\bar\omega/ck_z$, $s/ck_z$ and can be written in dimensionless form by redefining 
\be\label{eq:dimensionless}
s \rightarrow \tilde s\cdot{ck_z\over\bar\gamma_0^2}; \quad\quad \bar\omega \rightarrow \varpi\cdot{ck_z\over\bar\gamma_0^2};
\quad\quad \varepsilon \equiv  (k_{p,\rm ex}\Delta)^2{k_{p,\rm ex}^2\over 4(k_z^2-k_{p,\rm ex}^2)}.
\ee
The real and imaginary parts of Equation (\ref{eq:disp2}) become
\be
   \tilde s^4 - 6 \tilde s^2 \varpi^2 + \varpi^4 =
   4(1-2\varpi)\varepsilon;\quad\quad  \tilde s(\tilde s^2\varpi - \varpi^3 - 2\varepsilon) = 0.
\ee
Eliminating $\tilde s^2$ (we discard the case $\tilde s = 0$) gives
\be\label{eq:omeq}
\varpi^6 + \varepsilon\varpi^2 = \varepsilon^2;\quad\quad \tilde s = \left({2\varepsilon\over\varpi} + \varpi^2\right)^{1/2}.
\ee

We are interested in the real roots of the polynomial in Equation (\ref{eq:omeq}) that also yield a real solution for  $\tilde s$, and
are consistent with ${\rm Re}[\kappa_{\rm in}^2] < 0$.  The internal and external coefficients $\kappa^2$ can now be written as
\be\label{eq:kapin}
   \kappa_{\rm in}^2
   \;\simeq\; {\left(1-2\varpi -2i\tilde s\right)\over (\tilde s - i\varpi)^2}\, {k_{p,\rm ex}^2\over\bar\gamma_0};\quad\quad
   \kappa_{\rm ex}^2
   \;\simeq\; (1-2\varpi -2i\tilde s)\,{k_z^2 - k_{p,\rm ex}^2\over\bar\gamma_0^2}.\quad\quad
\ee
We then have, after substituting for $\tilde s^2$, 
\be
{\rm Re}[\kappa_{\rm in}^2] \propto {2\varepsilon(1+2\varpi) + 4\varpi^4\over \varpi}.
\ee
This quantity can be negative only if at least one of $\varpi$ and $\varepsilon$ is negative.
The case $\varepsilon > 0$ does not yield any solution with real $\tilde s$ and
${\rm Re}[\kappa_{\rm in}^2] < 0$.   Therefore we focus on the case $\varepsilon < 0$,
corresponding to $k_z < k_{p,\rm ex}$.

The cubic polynomial in Equation (\ref{eq:omeq}) has one positive root $\varpi^2$
when $\varepsilon < 0$.  Straightforward analytic solutions for
$\varpi$ and $\tilde s$ at large and small $|\varepsilon|$ are
\ba\label{eq:mode0}
\varpi &\;\simeq\;& -|\varepsilon|^{1/4}; \quad \tilde s \;\simeq\; |\varepsilon|^{1/4};  \quad\quad (|\varepsilon| \ll 1);\nn
\varpi &\;\simeq\;& -|\varepsilon|^{1/3}; \quad \tilde s \;\simeq\; \sqrt{3}|\varepsilon|^{1/3};  \quad\quad (|\varepsilon| \gg 1).
\ea
The positive solution for $\varpi$ can be discarded,
because it implies either a positive value of ${\rm Re}[\kappa_{\rm in}^2]$ or imaginary $\tilde s$.
Because $k_z < k_{p,\rm ex}$, we must remember that $|\varepsilon|$ is bounded below,
\be\label{eq:bound}
|\varepsilon| >  {(k_{p,\rm ex}\Delta)^2\over 4},
\ee
with this bound depending on the current sheet thickness.  
More general expressions for $\varpi$ are obtained from the standard solution to the cubic polynomial equation,
\ba\label{eq:mode1}
\varpi &=& -{2^{1/2}|\varepsilon|^{1/4}\over 3^{1/4}}
\cos^{1/2}\left[{1\over 3}\arccos\left({3^{3/2}|\varepsilon|^{1/2}\over 2}\right)\right]
\quad\quad \left(|\varepsilon| < {4\over 27}\right);\nn
&=& -\left({|\varepsilon|\over 3}\right)^{1/4}\left(X^{1/3} + X^{-1/3}\right)^{1/2}\quad\quad
\left(|\varepsilon| > {4\over 27}\right);\nn
   X &\equiv& \left(-1 + {27\over 4}|\varepsilon|\right)^{1/2} + \left({27\over 4}|\varepsilon|\right)^{1/2}.
\ea
Restoring dimensionful units, the phase speed of the mode lags the particle speed by the amount
\be\label{eq:mode3}
   {\omega\over k_z} - \bar\beta_0c \; = \; {\bar\omega\over k_z} \; \simeq \;  2\varpi(1-\bar\beta_0)c \;<\; 0.
\ee
When $|\varpi| = O(1)$ this is comparable to the lag of the particle speed with respect to the speed of light.   
The mode growth rate is obtained from
\be\label{eq:mode2}
{s\over ck_z} = {\tilde s \over \bar\gamma_0^2} = 
{1\over \bar\gamma_0^2}\, \left(\varpi^2 - {2|\varepsilon|\over\varpi}\right)^{1/2}.
\ee

\begin{figure}
  \epsscale{1.05}
  \plottwo{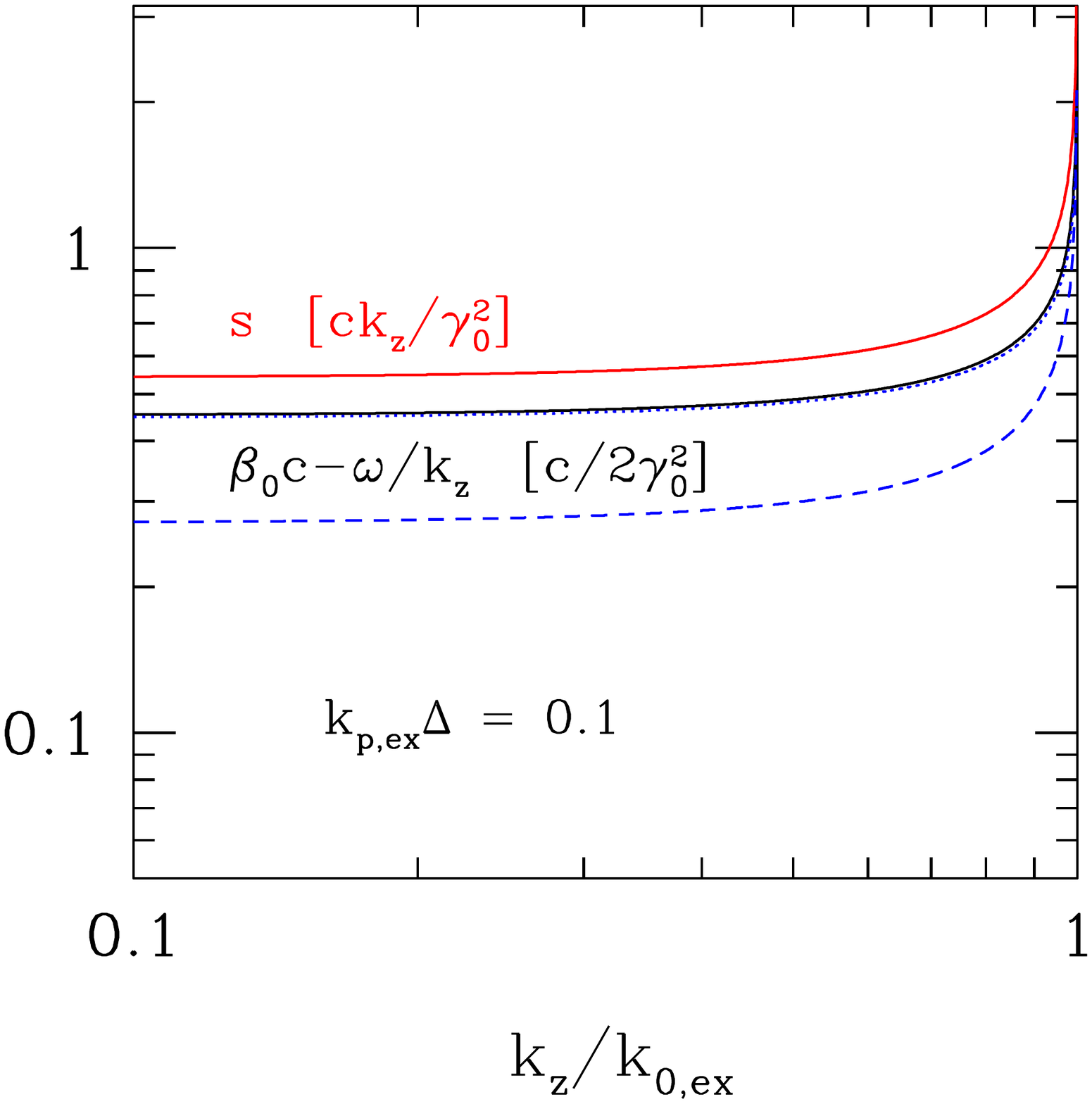}{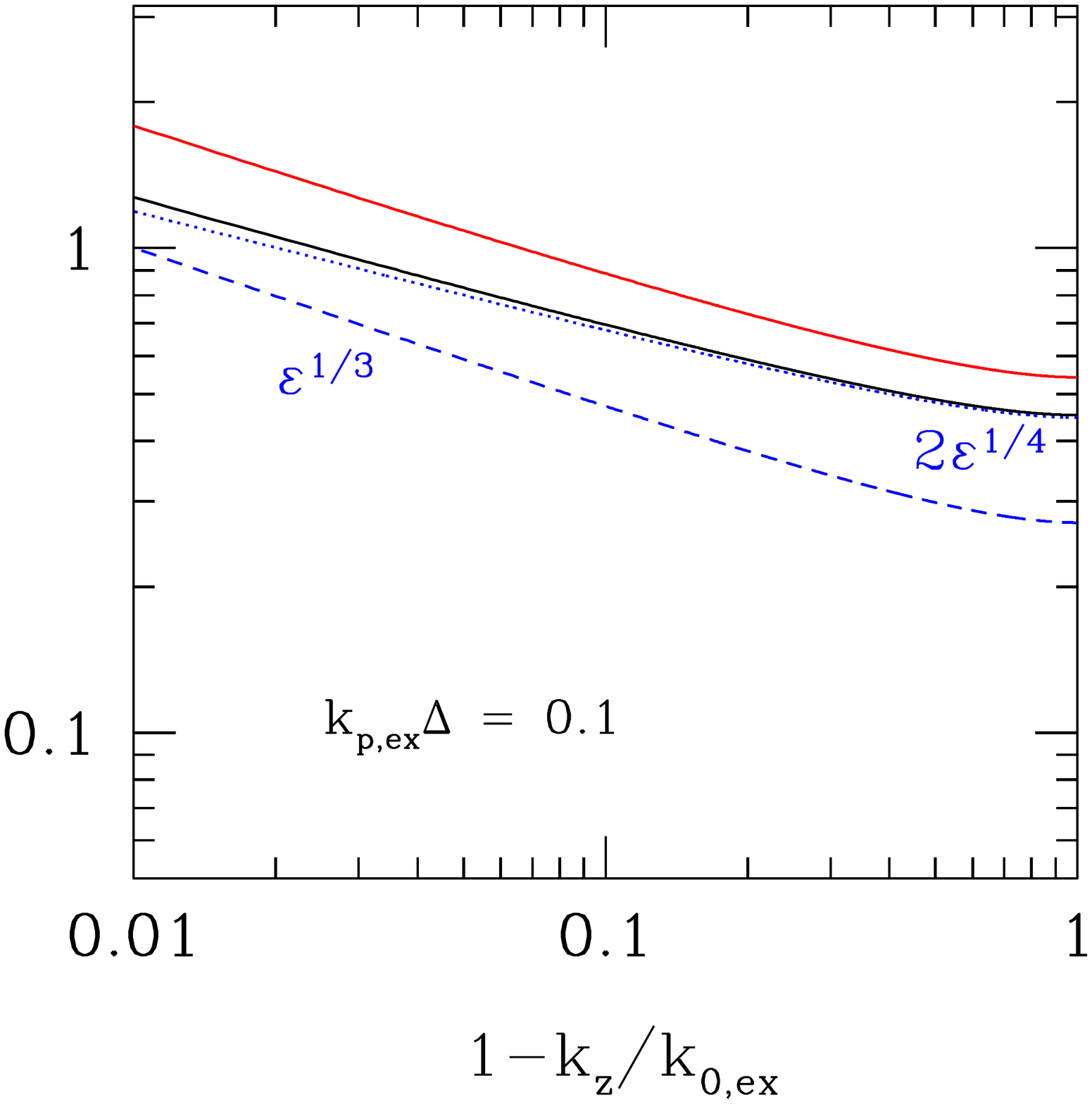}
  \vskip -0.7in
  \plottwo{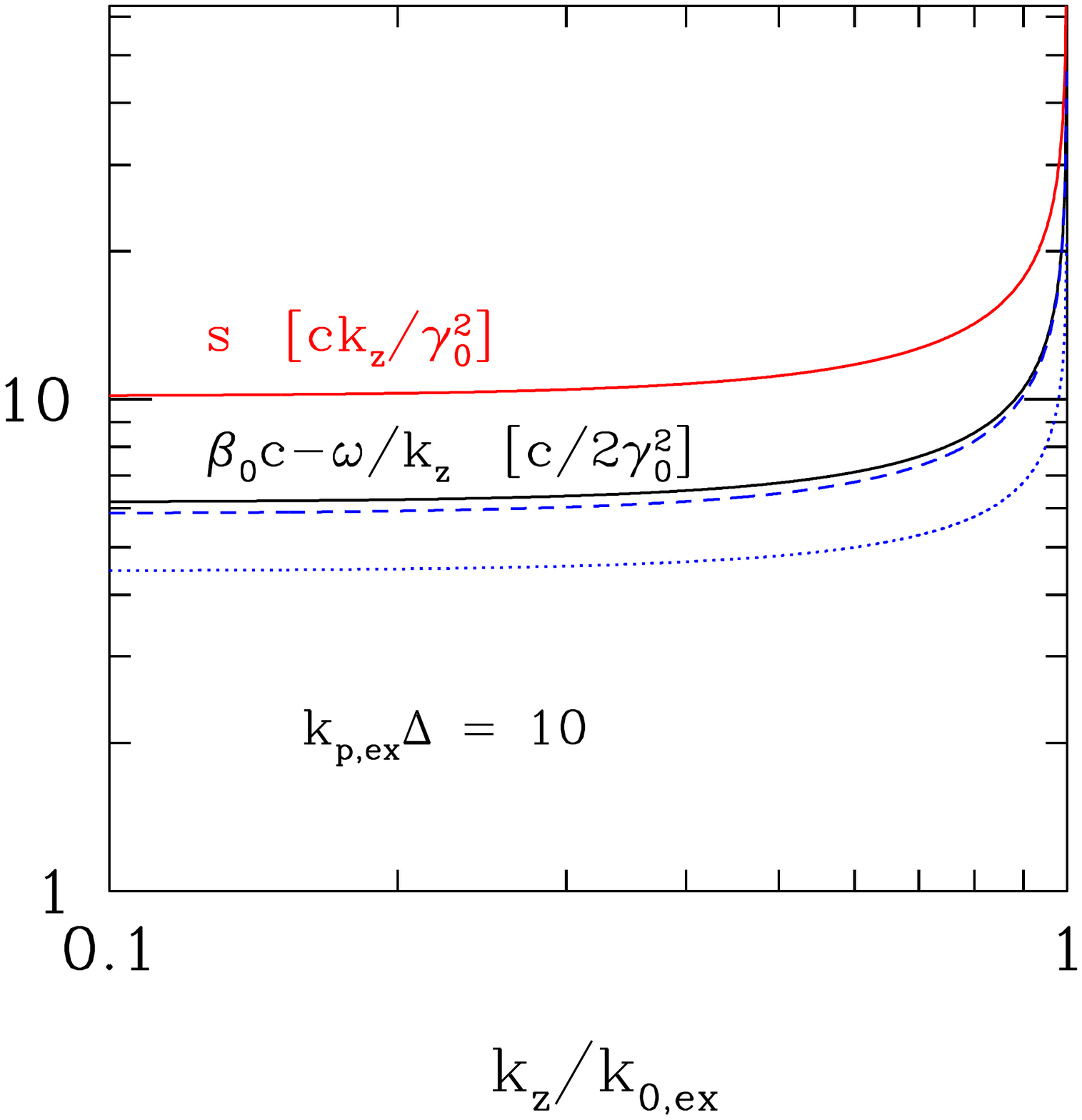}{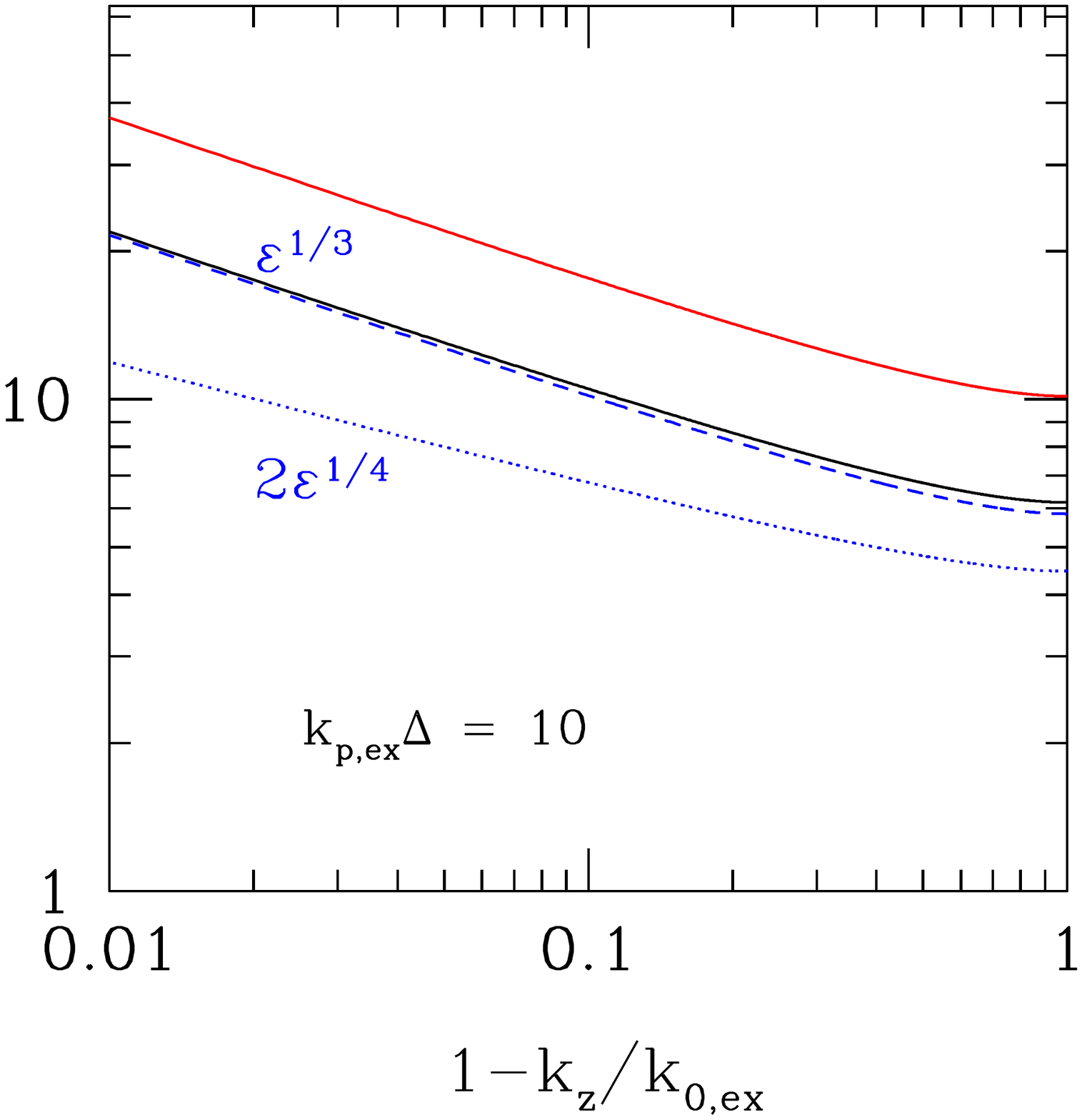}  
  \vskip -.5in
  \caption{Dispersion relation of sheared electromagnetic wave localized at a current sheet
    ($-\Delta < x < \Delta \sim k_{p,\rm ex}^{-1}$) in a quantizing magnetic field.
    Charges in the current sheet stream relativistically along the magnetic field with speed $\bar\beta_0 c$.
    Mode propagates parallel to the guide magnetic field with wavenumber $k_z$, and with vanishing wavenumber $k_y$
    in the orthogonal direction in the plane of the sheet.
    Overstable modes are found for $k_z < k_{p,\rm ex}$ ($\varepsilon < 0$, Equation (\ref{eq:dimensionless})).
    Lag of the phase speed relative to the particle speed (black curves) and mode
    growth rate (red curves) are plotted vs. $k_z/k_{p,\rm ex}$ (left panels) and
    $1-k_z/k_{p,\rm ex}$ (right panels, showing the fastest-growing part of the dispersion curve).
    Top panels:  thin current sheet ($k_{p,\rm ex}\Delta = 0.1$);  bottom panels:  thick current sheet ($k_{p,\rm ex}\Delta = 10$).
    Blue dotted curves show the low-$|\varepsilon|$
    approximation $\varpi = -|\varepsilon|^{1/4} = -(k_{p,\rm ex}\Delta/2)^{1/2}(1-k_z^2/k_{p,\rm ex}^2)^{-1/4}$,
    valid for small $|\varepsilon|$ -- see Equations (\ref{eq:dimensionless}) and (\ref{eq:mode0}).
    Blue dashed curves show the high-$|\varepsilon|$  approximation
    $\varpi = -|\varepsilon|^{1/3}$.\label{fig:dispersion}}
\end{figure}

\begin{figure}
  \epsscale{1.05}
  \vskip -0.3in
  \plottwo{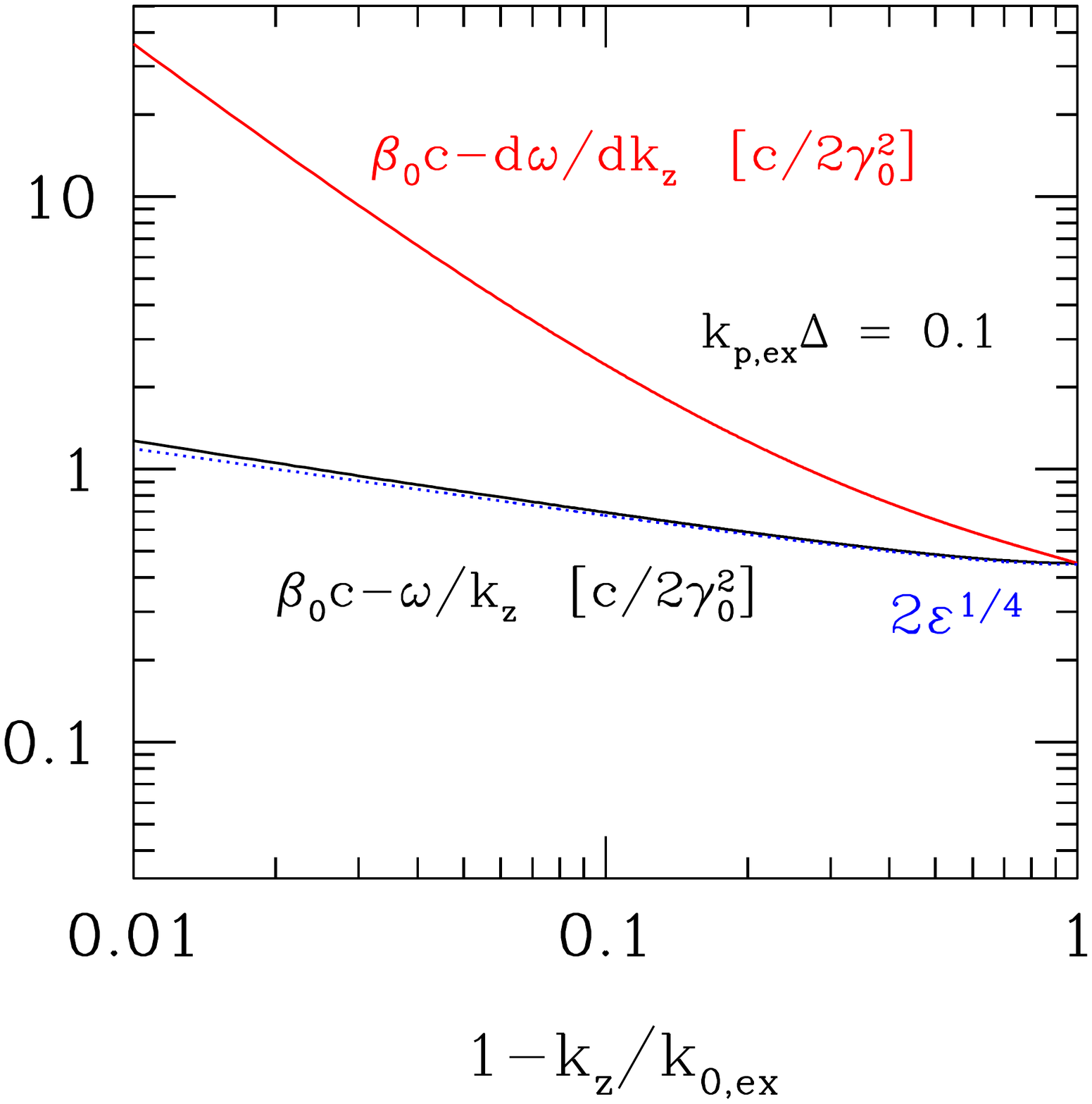}{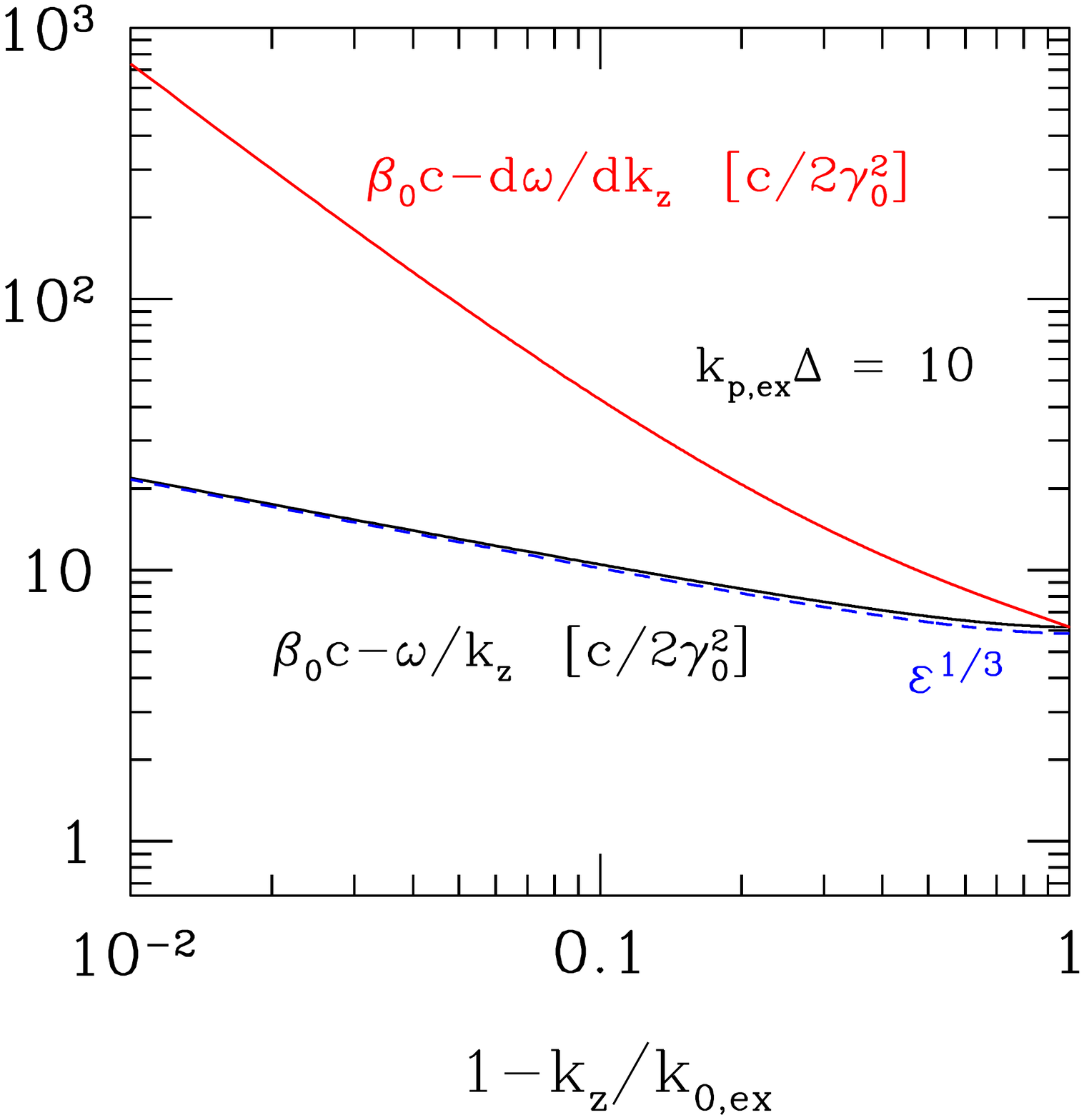}
  \vskip -.5in
  \caption{Right panel:  Comparison of the mode phase speed and group speed.  The mode propagates more
    slowly than the particles, with the group speed lagging the phase speed.\label{fig:velocity}}
\end{figure}

Figure \ref{fig:dispersion} shows the dispersion relation given by Equations (\ref{eq:mode1}) and (\ref{eq:mode2}),
for a thin current sheet ($k_{p,\rm ex}\Delta = 0.1$) and a thick sheet ($k_{p,\rm ex}\Delta = 10$).
The phase speed always lags the particle speed, $\bar\omega/k_z = \omega/k_z - \bar\beta_0c < 0$.
The dotted and dashed curves in Figure \ref{fig:dispersion} show that the approximate solution (\ref{eq:mode0})
is a good fit to the full dispersion relation both when $|\varepsilon| \ll 1$ (thin current sheet)
and $|\varepsilon| \gg 1$ (thick sheet).  

Two key properties of these growing modes are worth emphasizing.
\vskip .1in
\noindent 1. Fastest growth is obtained
when the real frequency is close to the ambient plasma frequency outside the current sheet, $\omega \simeq \omega_{p,\rm ex} = 
ck_{p,\rm ex}$.  This sets a lower bound to the frequency of the curvature radiation emitted by a charged wavepacket in a pulsar
magnetosphere (Section \ref{s:curve}).

\noindent 2. A mode with $k_z < k_{p,\rm ex}$ is localized about the current sheet only as long as $s \neq 0$.  Returning to
Equation (\ref{eq:kapin}), one sees that $\kappa_{\rm ex}$ is purely imaginary when $s = 0$.  The real part is
\ba\label{eq:realkap}
   {\rm Re}[\kappa_{\rm ex}] &\;=\;& {(k_{p,\rm ex}^2-k_z^2)^{1/2}\over\bar\gamma_0}
   \left[{2\varpi-1\over 2}+{1\over 2}\sqrt{(2\varpi-1)^2 + 4\tilde s^2}\right]^{1/2} \quad\quad (\varpi < 0)\nn
   &\;\sim\;& {(k_{p,\rm ex}^2-k_z^2)^{1/2}\over \bar\gamma_0}\tilde s\quad\quad (|\varpi|,\,\tilde s \ll 1).
\ea 
\vskip .1in

The peak growth rate of the mode is attained as $k_z \rightarrow k_{p,\rm ex}$.  However, the fastest growing modes
are not those most relevant for pulsar radio emission, because the group Lorentz factor $\bar\gamma_{\rm gr}$
also drops substantially below $\bar\gamma_0$,
thereby suppressing the peak curvature frequency (Section \ref{s:omcpk}).
For reference, a derivation of the peak growth rate is detailed in Appendix \ref{s:peak}.

We can also check the consistency of the approximation $|\kappa_{\rm in}|\Delta < 1$ that was used in the derivation of Equations
(\ref{eq:mode0}), (\ref{eq:mode1}), and (\ref{eq:mode2}).  One has
\be\label{eq:kapapprox}
|\kappa_{\rm in}^2| \Delta^2 \sim 
{(k_{p, \rm ex}\Delta)^2\over \bar\gamma_0}\,F(|\varepsilon|);
\quad\quad F(|\varepsilon|) \equiv {\rm max}\,\left[{1\over 2|\varepsilon|^{1/2}},\,{1\over |\varepsilon|^{1/3}}\right],
\ee
as can be seen by substituting the dispersion relation (\ref{eq:mode0})  in Equation (\ref{eq:kapin}).
Given that $|\varepsilon|$ has the lower bound (\ref{eq:bound}), we find
\be
|\kappa_{\rm in}|\Delta \;\ll\; 1  \quad \Leftrightarrow \quad \bar\gamma_0 \;\gg\; {\rm max}[(k_{p,\rm ex}\Delta)^{4/3},1].
\ee

\subsubsection{Group Speed}

The group speed of the mode
generally shows a stronger lag with respect to the particle speed $\bar\beta_0c$ than does its phase speed
(red curves in Figure \ref{fig:velocity}).  It is obtained by differentiating the polynomial in
Equation (\ref{eq:omeq}),
\be\label{eq:vgr}
   \beta_{\rm gr} c - \bar\beta_0c \;\equiv\; {d\omega\over dk_z} - \bar\beta_0c \;\simeq\;
    {c\over\bar\gamma_0^2}\left[\varpi - {\varepsilon^2 + \varpi^6\over
       \varpi(\varepsilon + 3\varpi^4)}{k_z^2\over k_z^2 - k_{p,\rm ex}^2}\right] < 0.
\ee
The group Lorentz factor $\gamma_{\rm gr} = (1-\beta_{\rm gr}^2)^{-1/2}$ plays an important role in the discussion
of curvature emission (Section \ref{s:curve}).
Its leading dependence on $|\varepsilon|$ is, from Equation (\ref{eq:vgr}),
\be\label{eq:gamgr}
   {\gamma_{\rm gr}\over \bar\gamma_0} \;\simeq\;
   \begin{cases}
     [1 + 16|\varepsilon|^{4/3}/ 3(k_{p,\rm ex}\Delta)^2]^{-1/2} \;\sim\; 1.1\,(k_{p,\rm ex}\Delta)^{-1/3}(1-k_z^2/k_{p,\rm ex}^2)^{2/3}
     & \quad\quad (|\varepsilon| \gg 1);\\
     [1 + 4|\varepsilon|^{5/4} / (k_{p,\rm ex}\Delta)^2]^{-1/2} \;\sim\;  1-0.35(k_{p,\rm ex}\Delta)^{1/2}(1-k_z^2/k_{p,\rm ex}^2)^{-5/4}
     & \quad\quad (|\varepsilon| \ll 1).\\
   \end{cases}
\ee
In general, a wavepacket propagates more slowly with increasing linear mode growth rate.
When the current sheet is thick ($|\varepsilon| \gg 1$), the mode propagates substantially more slowly
than the charged particles;  when the sheet is thin, the lag between group speed and particle speed is comparable to the
lag between the particles and the speed of light at a wavenumber $k_z^2/k_{p,\rm ex}^2 \sim 1-(k_{p,\rm ex}\Delta)^{1/2}$.

\subsubsection{Mode Saturation}\label{s:sat}

\begin{figure}
  \epsscale{0.55}
  \vskip -0.5in
  \plotone{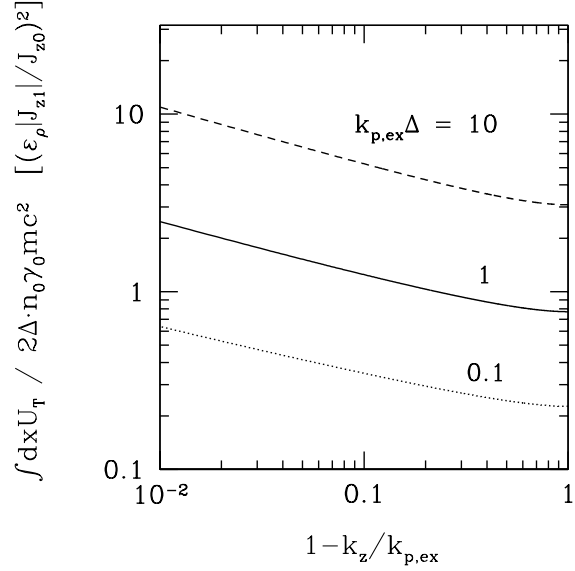}
  \vskip -0.5in
  \caption{Energy carried by the transverse electromagnetic component of the mode, as given by Equation
    (\ref{eq:Etrans}), normalized by the kinetic energy of the background particle flow and the nonlinearity
    parameter $\varepsilon_\rho|J_{z1}|/J_{z0}$ (Equation (\ref{eq:dJ})).  The result plotted
    is independent of $\bar\gamma_0$, under the assumption that $\bar\gamma_0 \gg 1$.\label{fig:Etrans}}
\end{figure}

The main constraint on the mode amplitude follows from a comparison of the wave energy and the kinetic energy of
the background particle flow.   Similarly to the
Alfv\'en wave propagating through a homogeneous, current-free plasma (Section \ref{s:alfven}), the 
wave energy can be divided into a transverse electromagnetic component, and a longitudinal component.
The transverse wave energy density is
\be
  U_T =  {|B_{y1}|^2 + |E_{x1}|^2\over 8\pi} = {|\partial_x A_{z1}|^2 + |\partial_x\phi_1|^2\over 8\pi} =
   \left(1 + {c^2k_z^2\over\omega^2}\right){|\partial_x A_{z1}|^2\over 8\pi}.
\ee
The longitudinal excitation divides into kinetic and electrostatic components, which as we now show carry less energy
than the transverse component of the wave.  A second-order kinetic energy perturbation
is sourced by the first-order parallel electric field perturbation $E_{\parallel,1}$  and
velocity perturbation $\beta_1 = qE_{\parallel,1}/\bar\gamma_0^3(\bar\omega + is) mc$,
\be
   {d\gamma_2\over dt} = {q\beta_1\over mc} E_{\parallel,1}.
\ee
The total longitudinal energy perturbation is 
\be\label{eq:Elong}
U_L = n_0 |\gamma_2| mc^2 + {|E_{\parallel,1}|^2\over 8\pi} =
\left[{k_{p,\rm in}^2c^2\over \bar\omega^2 + s^2} + 1\right]{|E_{\parallel,1}|^2\over 8\pi}.
\ee

The electromagnetic
field components have profiles $B_{y1}(x),\,E_{x1} \propto A_{z1}'(x)$ and $E_{\parallel1,}(x) \propto A_{z1}(x)$
in the coordinate $x$ running transverse to the current sheet.
Integrating the field energy over $x$ using the interior and exterior eigenfunctions (\ref{eq:Eparvsx}),
in the regime $|\kappa_{\rm in}|\Delta \ll1$ where the dispersion relation has been derived, gives
\ba\label{eq:xint}
{1\over 2\Delta}\int_{-\infty}^\infty dx |B_{y1}(x)|^2 &\;=\;& 
  |A_{z1}(0)|^2\,|\kappa_{\rm ex}|^2\left[{1\over 3} + {1\over 2{\rm Re}[\kappa_{\rm ex}])\Delta}\right];  \nn
  {1\over 2\Delta}\int_{-\infty}^\infty dx |E_{\parallel,1}(x)|^2 &\;=\;&  |E_{\parallel,1}(0)|^2
\left[1 + {1\over 2{\rm Re}[\kappa_{\rm ex}]\Delta}\right].
\ea
The two terms inside the brackets are the contributions from inside and outside the current sheet;
the decay coefficient ${\rm Re}[\kappa_{\rm ex}]$  of the perturbed field outside the sheet is given by Equation (\ref{eq:realkap}).
Normalizing the field energy to the kinetic energy of the charges streaming through the sheet gives
\be\label{eq:Etrans}
{\int_{-\infty}^\infty dx\, U_T\over 2\Delta\cdot n_0\bar\gamma_0mc^2} \;=\; {(k_{p,\rm ex}\Delta)^2\over \bar\gamma_0^3}\,
\left[{1\over 3} + {1\over 2{\rm Re}[\kappa_{\rm ex}])\Delta}\right] F(|\varepsilon|)^2\,|{\cal A}_{z1}(0)|^2
\ee
Here,
\be
   {\cal A}_{z1} \equiv {qA_{z1}\over mc^2}
\ee
is the normalized potential perturbation.   We have used Equations (\ref{eq:disp}) and (\ref{eq:kapapprox})
for $\kappa_{\rm in,\,ex}$.  The result is plotted in Figure \ref{fig:Etrans}. 

The longitudinal mode energy (\ref{eq:Elong}) is generally subdominant, except when peak growth rate is approached.
For simplicity, we show the component inside the current sheet;
the contribution from the exterior is obtained from Equation (\ref{eq:xint}).
The longitudinal electric field as given by
Equation (\ref{eq:Epar}) can be re-written as $E_{\parallel,1} \simeq -i (1-2\varpi -2i\tilde s)k_zA_{z1}/\bar\gamma_0^2$.  Hence, 
\be\label{eq:Elong2}
   {\int_{-\infty}^\infty dx\,|\gamma_2|\over 2\Delta\cdot \bar\gamma_0}
        = {(1-2\varpi)^2 + 4\tilde s^2\over 2(\varpi^2 + \tilde s^2)}{|{\cal A}_{z1}(0)|^2\over\bar\gamma_0^4};
\ee
\be
{\int_{-\infty}^\infty dx \, E_{\parallel,1}^2/8\pi \over 2\Delta\cdot n_0\bar\gamma_0 mc^2} = 
{k_z^2\over \bar\gamma_0 k_{p,\rm ex}^2}(\varpi^2 + \tilde s^2)\,
   {\int_{-\infty}^\infty dx\,|\gamma_2|\over 2\Delta\cdot \bar\gamma_0}.
\ee
Comparing with Equation (\ref{eq:Etrans}), one sees that the transverse wave carries a factor $\sim \bar\gamma_0$
more energy than the longitudinal kinetic energy perturbation when $|\varpi|,\;\tilde s = O(1)$.  Near
peak growth (where $|\varpi|,\;\tilde s \gg 1$; see Appendix \ref{s:peak}), the two components carry comparable energy.

The potential perturbation $A_{z1}$ inside the current sheet carries a charge and current density
\be\label{eq:rho1}
 \rho_1 = {k_z\over\omega + is} J_{z1} \simeq \pm {1\over c} J_{z1}; \quad\quad
           {4\pi  J_{z1}\over c} \;\simeq\; {k_{p,\rm ex}^2\over \bar\gamma_0}{(1-2\varpi - 2i\tilde s)\over (\varpi + i\tilde s)^2}A_{z1}
    \quad\quad (|x| < \Delta);
\ee
see Equations (\ref{eq:Jzin}) and (\ref{eq:kapin}).  The current density perturbation outside the sheet is weaker by a
factor  $\sim 1/\bar\gamma_0$,
\be\label{eq:rho2}
{4\pi J_{z1}\over c} \;\simeq\; {k_{p,\rm ex}^2\over\bar\gamma_0^2}(1-2\varpi - 2i\tilde s)A_{z1}
    \quad\quad (|x| > \Delta),
\ee
but can contribute significantly to the total charge advected by a wavepacket (Section \ref{s:charge}).
Substituting the asymptotic expressions (\ref{eq:mode0}) for $\varpi$ and $\tilde s$,
the current perturbation is, relative to the background,
\be\label{eq:dJ}
   {|J_{z1}|\over J_{z0}} =
   {{\cal A}_{z1}\over \varepsilon_\rho\bar\gamma_0}\,F(|\varepsilon|)\quad\quad (|x| < \Delta),
\ee
Hence Equation (\ref{eq:Etrans}) can be written as
\be
{\int_{-\infty}^\infty dx\, U_T\over 2\Delta\cdot n_0\bar\gamma_0mc^2} \;=\; {(k_{p,\rm ex}\Delta)^2\over \bar\gamma_0}\,
\left[{1\over 3} + {1\over 2{\rm Re}[\kappa_{\rm ex}])\Delta}\right]\left(\varepsilon_\rho{|J_{z1}|\over J_{z0}}\right)^2.
\ee

When choosing a criterion for mode saturation, the physical nature of the instability should be kept in mind.  Our kinetic
treatment shows that the overstable trapped Alfv\'en mode is driven by the differential streaming of charges between the
current sheet and the exterior zone.  The weak, transverse background magnetic field $B_{y0}$ does not enter into
the perturbation equations as long as $k_y/k_z \ll B_z/|B_{y0}|$.  Therefore the mode energy is limited by the kinetic
energy of the charges, not by the current perturbation $|J_{z1}|/J_{z0}$.  In particular, the mode growth does not
depend on a net charge asymmetry $\varepsilon_\rho$, even though such an asymmetry is guaranteed to be present in the
magnetosphere of a rotating neutron star.

For these reasons, we adopt a saturation condition corresponding to
$\int dx U_T  \sim 2\Delta\cdot \bar\gamma_0 n_0m_c^2$, or equivalently
\be
\varepsilon_\rho {|J_{z1}|\over J_{z0}} \sim (k_{p,\rm ex}\Delta)^{-1/2}.
\ee
Here, we have estimated ${\rm Re}[\kappa_{\rm ex}] \sim k_{p,\rm ex}/\bar\gamma_0$ from Equation (\ref{eq:realkap}).

\subsection{Mode Growth with Particle Flow Outside the Current Sheet}\label{s:streamext}

The mode dispersion relation is easy to obtain in a more general case where a uniform flow is present
outside the current sheet, with speed and Lorentz factor
$\bar\beta_{0,\rm ex} \lesssim \bar\beta_0$, $\bar\gamma_{0,\rm ex} \lesssim
\bar\gamma_0$.  This is, in many respects, the most realistic case.  The energy that could be deposited
per particle in the sheet by the decay of the transverse magnetic field $B_{y0}$ is
\be
   {B_{y0}^2\over 8\pi n_0} \sim {1\over 2} (k_{p,\rm ex}\Delta)^2\cdot m_ec^2.
\ee
when the medium outside the current sheet is at rest.  If the sheet thickness is comparable
to the external skin depth, then it is natural for the particles inside the current sheet
to gain a moderately relativistic motion with respect to the exterior.  An additional bulk relativistic
motion, superimposed on this differential motion, is needed
for radio curvature emission.  This is readily supplied by secondary pair creation by
the conversion of curvature gamma rays \citep{sturrock71}:  the density of the created pairs is
uniform on the scale of the skin depth.

The Alfv\'en mode growth rate
for $\bar\gamma_{0,\rm ex} > 1$ can be obtained by a Lorentz transformation to the rest frame of the medium
external to the current sheet.
Although the transverse magnetic field $B_{y0}$ inside the current sheet is not invariant under a Lorentz boost
in the $z$-direction parallel to the sheet, this field is so weak that it does not enter directly into the dispersion
relation:  we take ${\bf k}\cdot {\bf B}_0 \rightarrow k_zB_{0z}$.
The sheet thickness is now evaluated in terms of the external skin depth
$k'_{p,\rm ex}$ in this rest frame.  Other Lorentz-transformed quantities are
\be
\bar\gamma_0' \;=\; \bar\gamma_{0,\rm ex}(1-\bar\beta_{0,\rm ex}\bar\beta_0)\bar\gamma_0 \;\simeq\;
{\bar\gamma_0\over(1+\bar\beta_{0,\rm ex})\bar\gamma_{0,\rm ex}};\quad\quad
\gamma_{\rm gr}' \;\simeq\; {\gamma_{\rm gr}\over (1+\bar\beta_{0,\rm ex})\bar\gamma_{0,\rm ex}}.
\ee
The rest-frame growth rate
\be\label{eq:srest}
s' = \tilde s(k_z'/k_{p,\rm ex}', k'_{p,\rm ex}\Delta){ck_z'\over (\bar\gamma_0')^2},
\ee
where $\tilde s$ is the same function as appearing in Equation (\ref{eq:mode2}).

The corresponding modifications to the charge, peak curvature emission frequency, and radiation power
of a nonlinear wavepacket are addressed in Appendix \ref{s:streamext2}.

\section{Coherent Emission of Superluminal Electromagnetic Waves}\label{s:curve}

We now examine the radiative implications of the Cerenkov instability of trapped Alfv\'en waves
that was described in Section \ref{s:kinetic}.
At least two emission channels are available: (i) low-frequency
curvature radiation by charged Alfv\'en solitons and (ii)
maser emission triggered by the interaction of the primary electron-positron flow with the longitudinal
component of the Alfv\'en mode.  Curvature radiation shows more promise and is investigated in detail here,
for the reasons given in Sections \ref{s:flow} and \ref{s:pulsars}.

\subsection{General Constraints on Soliton Formation}

Efficient curvature emission has been argued to depend on the formation of charged solitons
  in the pulsar circuit \citep{gk71,mgp00,lmm18}.  The most popular approach to soliton formation
  is based on the rapid linear growth of a plasma mode. 
   The details of how quasi-linear waves may convert to charged solitons remain an active area
    of research:   \cite{mrm21a} question whether Langmuir solitons can form by a two-stream instability
    in a pair plasma.
      
The constraint of rapid mode growth
is a familiar one in the case of Langmuir waves;  it implies a tension between rapid growth
and curvature emission by Langmuir solitons above the plasma cutoff (e.g. \citealt{lmm18,mrm21a}).
The growth of trapped Alfv\'en waves is subject to a similar constraint:  starting with a conservative
growth criterion, we find that the peak of the curvature spectrum is $\sim 10^{-2}$ of the cutoff frequency.

We first briefly review the longitudinal instability.
A charged soliton that might form from the collision of two clouds of $e^\pm$ pairs will propagate
in the center-of-momentum frame of the collision, defined by Lorentz factor $\bar\gamma$.
The cloud size must exceed $\Delta \sim c/\omega_p$, where $\omega_p = \bar\gamma\omega_p'$
is the cutoff frequency in the frame of the star and $\omega_p'$ is the plasma frequency in the center-of-momentum frame.
The collision is completed over a distance $\sim \bar\gamma^2\Delta$
in the frame of the star.  Requiring this to be less than the decoupling length $\delta l_{\rm rad}
\sim R_c/\bar\gamma$ of curvature radiation from the soliton (due to its varying emission direction)
implies $\omega_p > \bar\gamma^3c/R_c$.  The peak of the curvature
spectrum is therefore concentrated below the plasma cutoff, and radiation into the ordinary mode (O-mode)
is restricted.   A similar result applies if the growth rate is evaluated in terms of a kinetic instability.

In the case of the Alfv\'en Cerenkov instability, the relevant transformation is to the
local rest frame of the particle flow outside the current sheet, by a Lorentz factor
$\bar\gamma_{0,\rm ex}$ (see Section \ref{s:streamext}).  In this primed
frame, only the particles inside the current sheet are in motion, with Lorentz factor $\bar\gamma_0'
\sim \bar\gamma_0/(1+\bar\beta_{0,\rm ex})\bar\gamma_{0,\rm ex}$
with respect to the exterior.  Equations (\ref{eq:mode0}) and
(\ref{eq:srest}) show that growth is most efficient in the primed frame, $s' \sim \omega_{p,\rm ex}'/(\bar\gamma_0')^2$,
when $\bar\gamma_0'$ is not much larger than unity.  This enhancement of the instability disappears following
Lorentz transformation back to the frame of the star, where the growth length is
$(1+\bar\beta_{0,\rm ex})\bar\gamma_{\rm 0,\rm ex}c/s' \sim
\bar\gamma_0^2c/(1+\bar\beta_{0,\rm ex})\omega_{p,\rm ex}$.  As before, we define the cutoff frequency in the
stellar frame as $\omega_{p,\rm ex} = \bar\gamma_{p,\rm ex} \omega_{p,\rm ex}'$.
The peak curvature frequency $\omega_c^{\rm peak} \sim 0.5\gamma_{\rm gr}^3c/R_c$ depends on the
group Lorentz factor of the soliton, which we estimate using Equation (\ref{eq:gamgr}) from
the linear analysis.  The suppression of $\omega_c^{\rm peak}$ compared with $\omega_{p,\rm ex}$
arises in significant part from the fact that $\gamma_{\rm gr} < \bar\gamma_0$. 

A few features of the Cerenkov instability are favorable for the formation of charged Alfv\'en solitons,
which we now note.
A full investigation of three-dimensional soliton structure is beyond the scope of this paper.
\vskip .1in
\noindent
1. A significant nonlinearity is present through the interaction of the
longitudinal wave degrees of freedom ($E_\parallel$ and $p$) with the background particle flow.  The transverse
magnetic perturbation, which is the main source of nonlinearity in the soliton solutions
of \cite{mikhail85} and \cite{spangler85}, is heavily suppressed compared with the guide (poloidal) field by a factor
$\sim (\Delta/R_{\rm cap})(R_{\rm cap}/r_{\rm NS})^3$, where $r_{\rm NS}$ is the stellar radius
and $R_{\rm cap} \simeq (\Omega r_{\rm NS}/c)^{1/2}r_{\rm NS}$ is the half-diameter of the open magnetic
field bundle.
\vskip .1in
\noindent
2. The growing mode is localized in one transverse cartesian direction -- 
a property which is not available to plane-polarized shear Alfv\'en waves.   In the latter case,
such a localization may appear possible when the description of the wave is restricted to
force-free electrodynamics \citep{tb98,gralla14}, but this property is lost when the
longitudinal dynamics of the charges is taken into account.  The localization of a linear Alfv\'en wave
near a current sheet is directly tied to a finite growth rate -- see Equation (\ref{eq:realkap}).
\vskip .1in
\noindent
3. Shear Alfv\'en wavepackets supporting a net displacement of the magnetic field
(as measured from the front to the back of the waveform) carry net electric charge.
Gauss' law implies that the longitudinal electric field $E_\parallel$ does not vanish on at least
one side of the wavepacket.  Nonetheless, the energy of the wave is dominated by the
transverse electromagnetic field (Section \ref{s:sat}).  In addition, only a limited number $N$
of wavelengths fit inside the radiation decoupling length, as derived below in Equation
(\ref{eq:Nwave}).  Therefore $E_\parallel$ may execute a constrained random walk over
a distance $\delta l_{\rm rad} = R_c/\gamma_{\rm gr}$ with a minimal cost in energy to the Alfv\'en waves.
A sequence of solitons will, in effect, radiate independently if their charges are Poisson distributed,
so that the net charge within a length $\delta l_{\rm rad}$ scales as $N^{1/2}$.
\vskip .1in
\noindent
4.  The linear mode structure calculated in Section \ref{s:kinetic}
and the charge estimated in Section \ref{s:charge} self-consistently
include the effects of Debye screening.  This suggests that the output in low-frequency extraordinary waves
is not strongly limited by additional screening effects, as \cite{glm04} have argued to be the case for Langmuir solitons.
The linear electromagnetic wave penetrates a substantial distance ($\sim \bar\gamma_0'/k_{p,\rm ex}'$) outside
the supporting current sheet.  The case for weak screening is strongest for an ensemble of closely packed
but differentially propagating solitons.
\vskip .1in
\noindent
5. Linear dispersal by differential propagation along the sheared magnetic field in the current sheet
is insignificant (Section \ref{s:lindisp}).
\vskip .1in
\noindent
Finally, we note the possible role of the radiation reaction force in inducing the spontaneous clumping of charged
particles moving on a curved trajectory \citep{gk71,asseo83,stupatov02};
in the most recent analyses, this process has been found to be ineffective \citep{kl10}.  The radiation reaction could
possibly combine with the the Cerenkov instability described here to enhance soliton formation.

\subsection{Finite Size of the Wavepacket and Limiting Peak Curvature Frequency}\label{s:size}

We now review the radiation of superluminal electromagnetic waves by charged solitons
as they are guided adiabatically along a curved magnetic field.   The power and frequency
of the emitted radiation are regulated by (i) the finite size of the wavepacket \citep{schwinger49,gk71}; and (ii) the group
Lorentz factor $\gamma_{\rm gr}$, which is less than but comparable to the kinetic Lorentz factor $\bar\gamma_0$ of the
charge flow along the magnetic field (Equation (\ref{eq:gamgr})).  The wavepacket is several orders of magnitude smaller
than the radius of curvature $R_c$ of the magnetic field, so that its group velocity is aligned with the local direction
of the magnetic field. 

Consider first a soliton of a very small size, carrying a charge $Q$.
It emits electromagnetic waves in two orthogonal polarization
modes, (i) with the electric vector perpendicular to the plane of curvature ($\perp$ or X mode) and
(ii) the electric vector lying in the plane of curvature ($\parallel$ or O mode)
\be\label{eq:pc1}
(P_c^\perp,\;P_c^\parallel) = \left({1\over 8},\;{7\over 8}\right){2\gamma_{\rm gr}^4\over 3} {Q^2c\over R_c^2}
\ee
\citep{jackson98}.  The curvature radiation is beamed within an angle $\sim 1/\gamma_{\rm gr}$ about the local
direction of the magnetic field.  The electromagnetic pulse decouples from the charge when the emission
direction shifts through this angle, corresponding to a displacement $\sim R_c/\gamma_{\rm gr}$ of the soliton
along ${\bf B}_0$.

The finite size $\ell_z$ of the soliton modifies the spectrum and power of the emitted electromagnetic wave
when the soliton subtends an angle $\theta = \ell_z/R_c > 1/\gamma_{\rm gr}$ with respect to the local center
of field curvature.  More specifically, the total radiation power shifts to \citep{schwinger49}
\be\label{eq:pc2}
P_c = \left({\sqrt{3}\over \theta}\right)^{4/3}\,{Q^2c\over R_c^2} \equiv {2\gamma_{\rm geom}^4\over 3} {Q^2c\over R_c^2}.
\ee
We may therefore define an effective Lorentz factor for a radiating, charged wavepacket
by equating $P_c^\perp + P_c^\parallel$ from Equation (\ref{eq:pc1}) with Equation (\ref{eq:pc2}),
\be\label{eq:gameff}
\gamma_{\rm eff} = {\rm min}\left(\gamma_{\rm gr}, \gamma_{\rm geom}\right) =
  {\rm min}\left[\gamma_{\rm gr}, \;\left({3\over 2}\right)^{1/4}\left({\sqrt{3}R_c\over \ell_z}\right)^{1/3}\right].
\ee
This defines both the beaming angle and peak frequency of the emitted curvature radiation.
We note that, because $\ell_z \sim 1/k_{p,\rm ex}$, the geometric Lorentz factor
\be\label{eq:gamgeom}
\gamma_{\rm geom} \sim (k_{p,\rm ex}R_c)^{1/3} = 100 \left({k_{p,\rm ex}^{-1}\over 10~{\rm cm}}\right)^{-1/3}
\left({R_c\over 100~{\rm km}}\right)^{1/3}
\ee
depends on the plasma frequency in the medium outside the current sheet,
but not directly on the Lorentz factor of the particle flow within the sheet.

The peak curvature frequency is determined by $\gamma_{\rm eff}$.  
When $\gamma_{\rm gr} < \gamma_{\rm geom}$, the soliton behaves like a point emitter and the energy spectrum
$\omega\cdot d^2E_c/d\omega dt$ peaks at \citep{jackson98}
\be\label{eq:omcpk}
(\omega_c^{\rm peak,\perp},\;\omega_c^{\rm peak\parallel}) \sim (0.47,\;0.69)\,
         {\gamma_{\rm gr}^3c\over R_c} \;=\; (1.1,\;1.6)\left({\gamma_{\rm gr}\over \gamma_{\rm geom}}\right)^3 {c\over l_z}
         \quad\quad (\gamma_{\rm gr} < \gamma_{\rm geom}).
\ee
Given that $\ell_z \sim 1/k_{p,\rm ex}$, we see that the curvature frequency is comparable to the ambient plasma frequency
when $\gamma_{\rm gr} > \gamma_{\rm geom}$,
\be
\omega_c^{\rm peak} \sim ck_{p,\rm ex}\quad\quad (\gamma_{\rm gr} > \gamma_{\rm geom}).
\ee

Whether the wavepacket emits preferentially as a point-like source of curvature radiation, or as a source of finite size,
depends on its ability
to grow.  As is shown in Section \ref{s:omcpk}, the rapid linear growth of
trapped Alfv\'en waves requires $\gamma_{\rm gr} < \gamma_{\rm geom}$, meaning that large-amplitude Alfv\'en wavepackets
will effectively be point-like emitters.  This also implies that emission in the $\parallel$ mode
(the O mode) is suppressed \citep{glm04}, and the electromagnetic output is suppressed by at least a
factor $\sim {1\over 8}$ compared with the vacuum curvature formula (Section \ref{s:power}).

\begin{figure}
  \epsscale{1.1}
  \vskip -0.3in
\plottwo{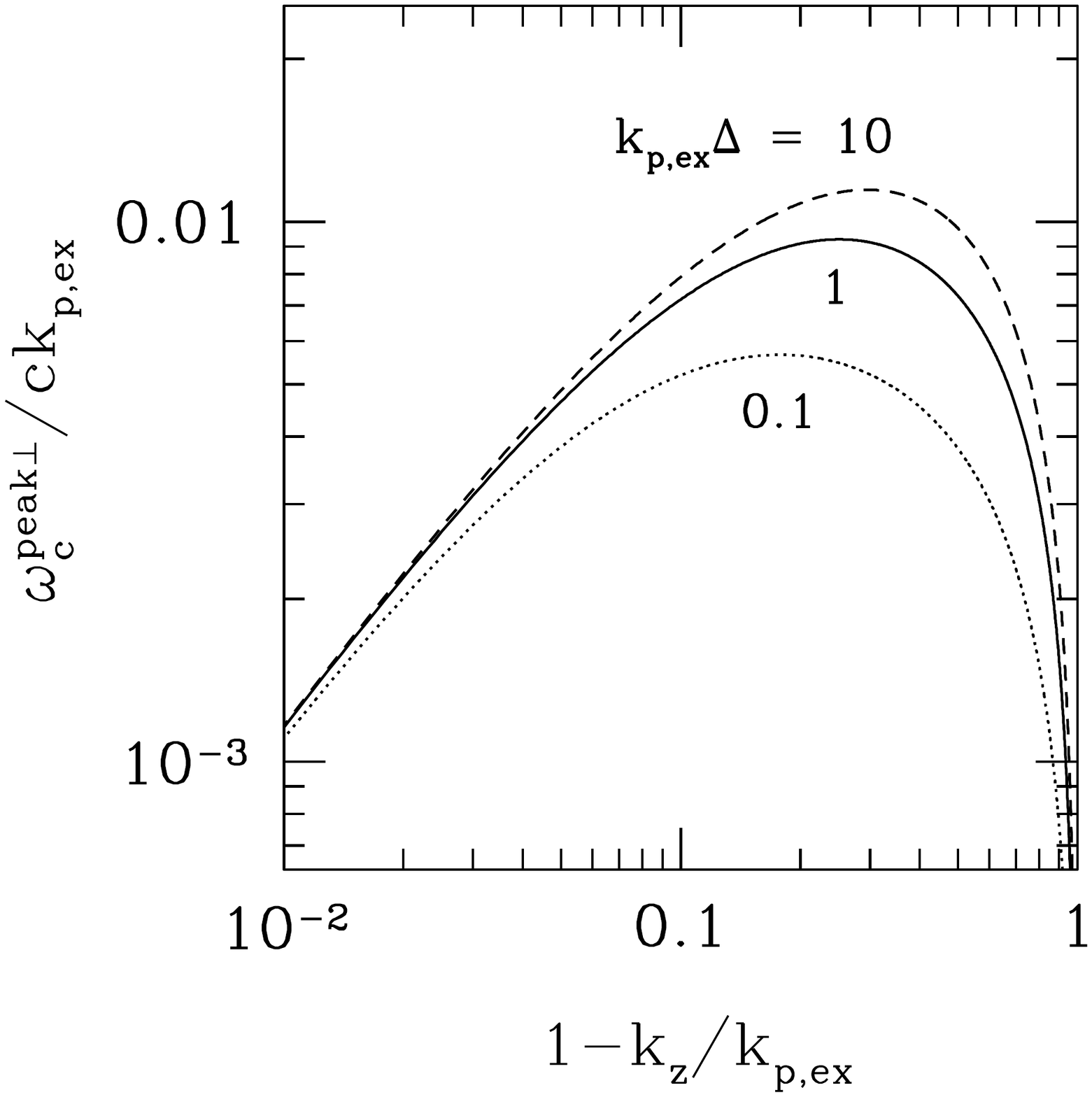}{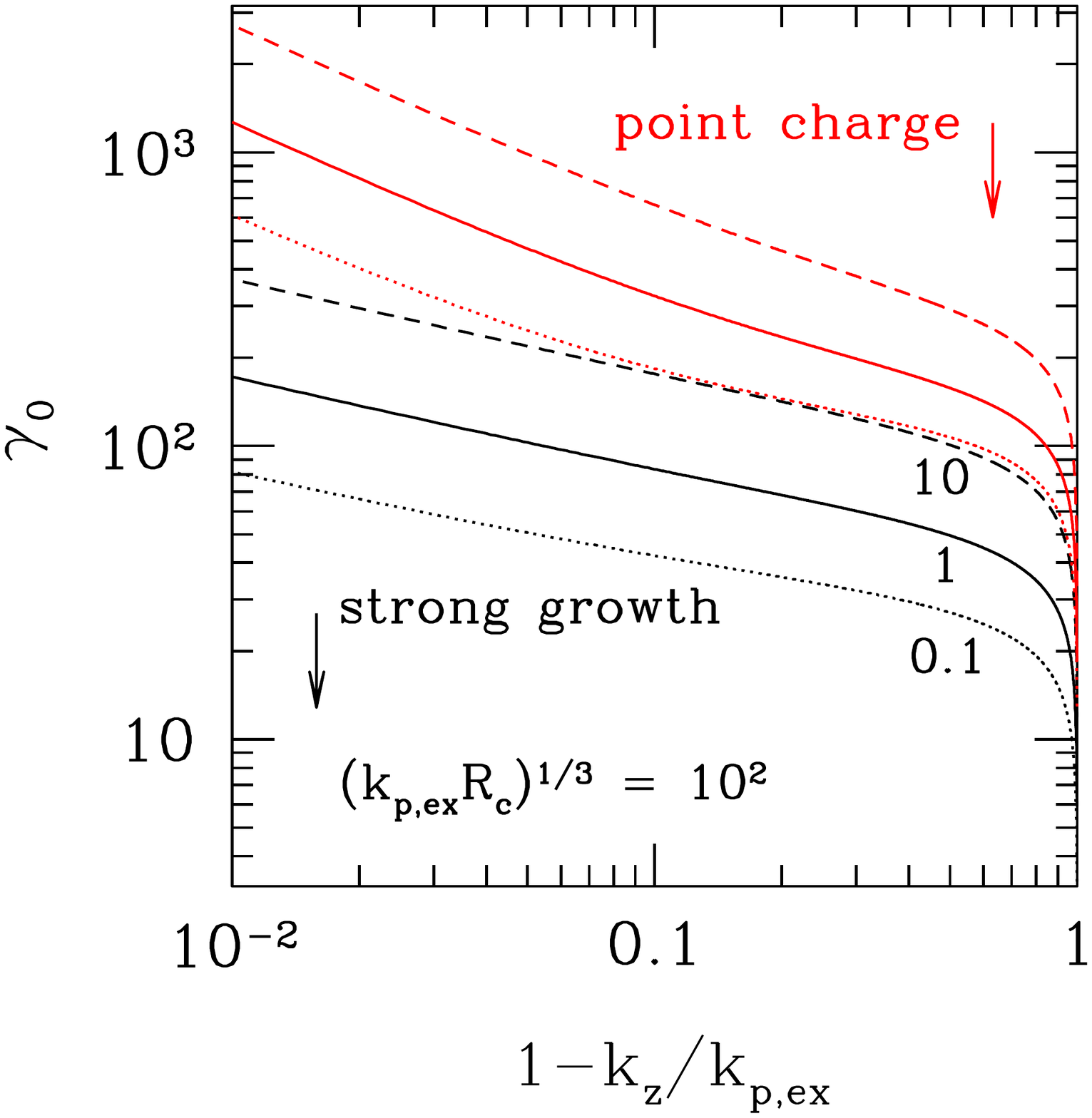}
  \vskip -.5in
  \caption{Left panel:  Peak frequency of X-mode curvature emission (electric vector perpendicular to the plane of curvature)
    as a function of the longitudinal wavenumber $k_z$.  Here, the group Lorentz factor of the Alfv\'en
      wavepacket is normalized to $\gamma_{\rm gr}^{\rm lin}$ of the linear mode (Equation (\ref{eq:vgr}));
      more generally $\omega_c \propto (\gamma_{\rm gr}/\gamma_{\rm gr}^{\rm lin})^3$.
    Right panel:  Bulk Lorentz factor of the charges flowing through the current sheet.  Black curve: Upper bound
    on $\bar\gamma_0$, corresponding to strong growth of the wavepacket according to Equation (\ref{eq:smin}).
    Red curves: Upper bound for radiation as a point charge,
    corresponding to $\gamma_{\rm gr} < \gamma_{\rm geom}$ (Equation (\ref{eq:gamgeom})).\label{fig:omcpk}}
\end{figure}

\subsection{Constraints on Curvature Radiation from Rapid Mode Growth}\label{s:omcpk}

The linear Alfv\'en mode must grow at a minimal rate if the radiation emitted by a soliton
is to track local plasma conditions.  The growth length $c/s$
must be shorter than the radiation decoupling length $R_c/\gamma_{\rm eff}$.  We adopt the working criterion,
\be\label{eq:smin}
s \gtrsim 10 \gamma_{\rm eff} {c\over R_c} \sim {20\,\omega_c^{\rm peak\perp}\over\gamma_{\rm eff}^2}.
\ee
The effective emission Lorentz factor $\gamma_{\rm eff}$ (Equation (\ref{eq:gameff}))
is the minimum of $\gamma_{\rm gr}$ and the critical Lorentz factor $\gamma_{\rm geom}$ below which the
soliton radiates like a point charge.
Curvature emission is primarily in the X mode when $\gamma_{\rm gr} < \gamma_{\rm geom}$, the peak emission frequency
being suppressed below the ambient plasma frequency to  $\omega_c^{\rm peak} \sim (\gamma_{\rm gr}/\gamma_{\rm geom})^3ck_z$
(see Equation (\ref{eq:omcpk})).

Let us now determine the maximum value of $\omega_c^{\rm peak\perp}$ consistent with the strong growth criterion
(\ref{eq:smin}).  We focus
on the open dipolar magnetic field lines of a pulsar, at a distance $r \sim 100$ km from the star, where the curvature
radius $R_c \sim (rc/\Omega)^{1/2}$.  Substituting Equation (\ref{eq:smin}) into Equation (\ref{eq:omcpk})
and normalizing $\gamma_{\rm gr}$ to the group Lorentz factor (\ref{eq:gamgr}) of the linear mode,
we find
\be\label{eq:omcpk2}
         \omega_c^{\rm peak\perp}(k_z/k_{p,\rm ex}, k_{p,\rm ex}\Delta) \;\sim\; 0.05\; \tilde s {k_z\over k_{p,\rm ex}}
         \left({\gamma_{\rm gr}\over \bar\gamma_0}\right)^2\,ck_{p,\rm ex}.
\ee
The result is plotted in Figure \ref{fig:omcpk} for a range of current sheet thickness.
We see that $\omega_c^{\rm peak\perp}$ tops out at about $0.01\,ck_{p,\rm ex}$ when $k_z \sim 0.7\,k_{p,\rm ex}$.
This confirms that curvature emission of O-mode photons is strongly screened.   At this wavenumber, one can
confirm that $\gamma_{\rm gr} < \gamma_{\rm geom}$ (see the red curves in Figure \ref{fig:omcpk}).

Although the bulk Lorentz factor $\bar\gamma_0$ scales out of Equation (\ref{eq:omcpk2}), it is bounded by the requirement
of marginal growth (Equation (\ref{eq:smin})),
\be\label{eq:gam0max}
\bar\gamma_0 < 
\left({\bar\gamma_0\over\gamma_{\rm gr}}{k_z\over k_{p,\rm ex}}{\tilde s\over 10}\right)^{1/3}\,(k_{p,\rm ex}R_c)^{1/3}.
\ee
This is plotted in Figure \ref{fig:omcpk} for $k_{p,\rm ex}R_c \sim 0.1~{\rm cm}^{-1}\cdot 10^7~{\rm cm} \sim 10^6$,
corresponding to $\gamma_{\rm geom} \sim 10^2$.   The limiting value of $\bar\gamma_0$ is
in the range expected for secondary pairs produced by an electromagnetic cascade in the polar gap of a radio
pulsar \citep{hibschman01,th19}.

\subsection{Linear Dispersal of the Trapped Alfv\'en Mode}\label{s:lindisp}

A wavepacket will experience some dispersal due to a differential displacement $\Delta y$ in the direction
of the non-potential magnetic field $B_{y0}(x)$, which varies across the current sheet.
To show that this effect is small, we first note that dispersion relation derived in
Section \ref{s:overstable} does not depend on the form of $B_{y0}$, being sensitive only
to the profile of the particle flow.   The longitudinal wavenumber 
$k_\parallel = (k_z B_{z0} + k_y B_{y0})/|{\bf B}_0|$ enters the kinetic equations and is well approximated
by $k_\parallel \simeq k_z B_{z0}/|{\bf B}_0| \simeq k_z$ because $|B_{y0}| \ll B_{z0}$.

The magnitude of $B_{y0}$ can be expressed in terms of the current density.  Taking $J_{z0} \sim \rho_{\rm co}c$, 
\be
   {c\over 4\pi}{|B_{y0}|\over\Delta} \sim {\Omega B_{z0}\over 2\pi}.
\ee   
It should be emphasized that the tearing process described in Paper I is driven by a small-scale component
of the current.  Then
\be\label{eq:ByBz}
   {B_{y0}\over B_{z0}} \;\sim\; {2\Omega\Delta\over c} \;\sim\; {2\Omega\over ck_{p,\rm ex}} 
     \;=\; 4\times 10^{-8}\left({k_{p,\rm ex}^{-1}\over 10~{\rm cm}}\right) \left({P\over 0.1~{\rm s}}\right)^{-1}.
\ee
Here, $P = 2\pi/\Omega$ is the rotation period of the neutron star.

As the wavepacket propagates a distance $\Delta z \sim \delta l_{\rm rad} = R_c/\gamma_{\rm gr}$
(the radiation decoupling length), it experiences a weak distortion if $\Delta y$ is much smaller than its
transverse size, $k_y^{-1}$.  Hence, we require
\be
\Delta y \sim {B_{y0}\over B_{z0}}\Delta z \ll k_y^{-1};\quad\quad \Delta z \sim {R_c\over\bar\gamma_0}.
\ee
Substituting Equation (\ref{eq:omcpk}) for $R_c$ and Equation (\ref{eq:ByBz}), this becomes
\be
\gamma_{\rm gr}^2 \sim 10^4 \ll {\omega_c\over\Omega}\cdot{k_{p,\rm ex}\over k_y}.
\ee
This inequality is easily satisfied.

\begin{figure}
  \epsscale{0.55}
  \vskip -0.3in
\plotone{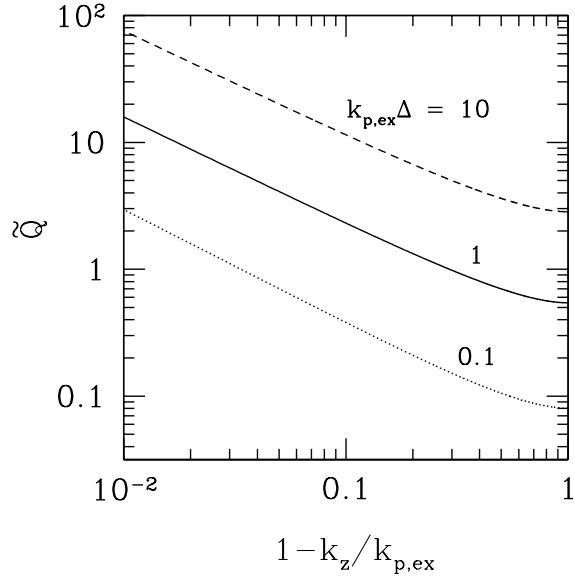}
  \vskip -.5in
  \caption{Dimensionless wavepacket charge as defined in Equation (\ref{eq:Qdef}).  A significant proportion
    of the charge is tied to the electromagnetic perturbation outside the current sheet, which decays away from the
    sheet over the lengthscale $1/{\rm Re}[\kappa_{\rm ex}]$ given by Equation (\ref{eq:realkap}).  The charge integral
    includes self-consistently the effects of Debye screening outside the current sheet.  Results are plotted for three
    values of the sheet thickness.\label{fig:charge}}
\end{figure}

\begin{figure}
  \epsscale{0.55}
  \vskip -0.3in
\plotone{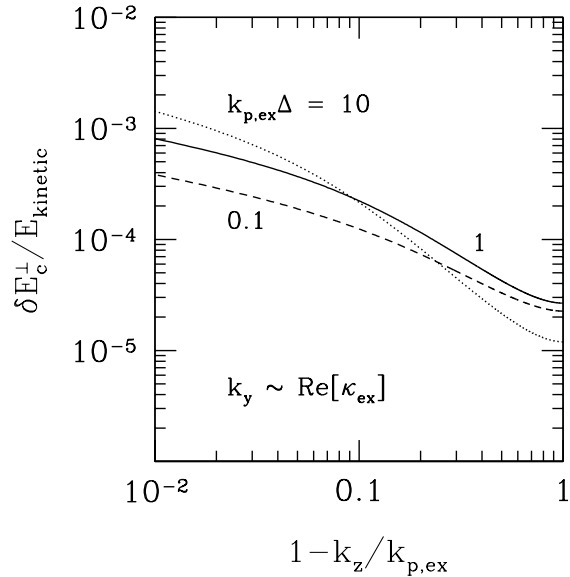}
  \vskip -.5in
  \caption{Energy in curvature radiation emitted by a charged wavepacket
    over the radiation decoupling length $\sim R_c/\gamma_{\rm eff}$, as a function of longitudinal
    wavenumber $k_z$ and normalized by the kinetic
    energy of the charges (Equation (\ref{eq:Eradnorm})).
    Group Lorentz factor of the packet is estimated using the linear mode dispersion relation, and the
    transverse packet size $k_y^{-1}$ parallel to the current sheet
    the penetration depth ${\rm Re}[\kappa_{\rm ex}]^{-1}$ of the trapped Alfv\'en mode into the plasma
    outside the sheet (Equation (\ref{eq:realkap})).   Result is shown for different values of the current sheet thickness.
    Net energy radiated over a length $\sim r$ is larger by a factor $10-10^2$.\label{fig:power}}
\end{figure}

\subsection{Efficiency of Curvature Emission}\label{s:power}

  Here we make a first estimate of the output in curvature radiation by a
  cluster of Alfv\'en solitons;  this provides a reference point for 
  numerical (e.g. PIC-based) explorations of plasma effects in the radiation process.  The size,
  group speed and charge of a single wavepacket are estimated using the saturated linear mode, e.g.,
    the linear mode evaluated at an amplitude corresponding to equality of the wave energy flux
    and the background particle energy flux.  The net output
  in curvature radiation depends on the clustering of solitons within a radiation wavelength
  (a feature of soliton radiation which is also open in the context of Langmuir-based models).
  We make the simplest assumption of a Poisson distribution of soliton charges.
  We obtain a relatively low emission efficiency as a result.

\subsubsection{Effective Radiating Charge}\label{s:charge}

We now estimate the maximum
effective radiating charge carried by a nonlinear Alfv\'en wavepacket, which will be used in Section
\ref{s:power} to estimate the energy transferred to curvature radiation.  
    Here, there are two qualitative differences with Langmuir waves: 
   first, the transverse electromagnetic field carries most of the mode energy and, second,
   the excited charge density field extends well beyond the
   driving particle flow.  
     The approach in this and the following sections
is simplified by the neglect of bulk relativistic motion in the exterior of a current sheet; the results are
easily generalized (Appendix \ref{s:streamext2}).

The power radiated at frequency $\omega$ by a current distribution ${\bf J}({\bf x}, t)$
in a direction ${\bf n}$ depends on integrals such as
$|\int d^3x e^{-i(\omega{\bf n}\cdot{\bf x}/c)}{\bf J}|^2$.  We will be interested in emission frequencies much
smaller than $\omega_{p,\rm ex} = ck_{p,\rm ex}$.  The effective charge carried by a wavepacket of
longitudinal size $k_z^{-1}$, transverse size $k_y^{-1}$, and charge density $\rho_1 \simeq J_{z1}/c$ is therefore taken
to be
\be
Q^2 \sim {1\over (ck_yk_z)^2}\int_{-\infty}^\infty dx J_{z1}(x)\,\int_{-\infty}^\infty dx [J_{z1}(x)]^*.
\ee
The linear mode used to compute this integral has been determined by a kinetic method that
includes screening effects.  The dominant contribution to the integral comes from
the medium outside the guiding current sheet:  the electromagnetic perturbation decays over a lengthscale
$1/{\rm Re}[\kappa_{\rm em}]$ (Equation (\ref{eq:realkap})) that is larger than the external skin depth $k_{p,\rm ex}^{-1}$
by a factor $\sim 1/\bar\gamma_0$.  
Combining Equations (\ref{eq:rho1}) and (\ref{eq:rho2}) for the current density with the mode profiles (\ref{eq:Eparvsx}),
and concentrating on the contribution to the $x$-integrals outside the current sheet, we obtain
\be
Q^2 \;\sim\;  {\bigl[(1-2\varpi)^2 + 4\tilde s^2\bigr]\over (k_yk_z)^2|\kappa_{\rm ex}|^2}
{k_{p,\rm ex}^4 |A_{z1}(0)|^2\over (4\pi \bar\gamma_0^2)^2}.
\ee 
Substituting further for $\kappa_{\rm ex}$ using Equation (\ref{eq:kapin}) and $A_{z1}$ in terms of the current
perturbation (\ref{eq:dJ}), we find
\be\label{eq:Qdef}
Q =  {qn_0 \over k_{p,\rm ex}k_yk_z} \left(\varepsilon_\rho{|J_{z1}|\over J_{z0}}\right)\,
\widetilde Q(k_z/k_{p,\rm ex},k_{p,\rm ex}\Delta),
\ee
where
\be\label{eq:tildeQ}
\widetilde Q(k_z/k_{p,\rm ex},k_{p,\rm ex}\Delta) = {\varpi^2 + \tilde s^2\over \left[(1-2\varpi)^2 + 4\tilde s^2\right]^{1/4}}
\left(1 - {k_z^2\over k_{p,\rm ex}^2}\right)^{-1/2}.
\ee
The result is shown in Figure \ref{fig:charge}  as a function of longitudinal wavenumber $k_z$ for different values of
the current sheet thickness.

\subsubsection{Net Radiative Output}

A few considerations arise here.   First, we have determined
that the peak curvature frequency lies a factor $\sim 10^{-2}$ below the plasma cut-off, leaving at most $\sim 1/8$
of the point-source curvature power available for emission in the X mode.  The energy radiated by an
isolated wavepacket of charge $Q$ propagating a distance $\sim R_c/\gamma_{\rm eff}$ along the curved magnetic field is
limited to
\be\label{eq:Ecperp}
\delta E_c^\perp \sim \left({1\over 8}\right){2\gamma_{\rm eff}^3\over 3}{Q^2\over R_c}
\sim {\omega_c^{\rm peak\perp}\over 6c}Q^2.
\ee

Second, one must consider whether distinct solitons propagating along the same magnetic field bundle
will radiate independently.  Their emission at frequency $\omega_c$ will add constructively or destructively
(depending on the relative signs of the charges) if, from Equations (\ref{eq:pc2}) and (\ref{eq:omcpk}),
the solitons are more closely spaced than a distance $\sim (\sqrt{3}/2)c/\omega_c$.

Third, the solitons receive a systematic charge of the same sign as the corotation charge, but of a small magnitude
when the pair multiplicity ${\cal M}_\pm = |e|n_+/|\rho_{\rm co}|$ is very large in the pulsar magnetosphere.  
The magnitude of this systematic charge is only a fraction $\varepsilon_\rho \sim 1/2{\cal M}_\pm$ of
the total soliton charge, which has either sign with equal probability.

It is at this stage that the relative motion of the charges inside and outside the current sheet becomes an
important consideration.  The number of independent solitons inside the radiation length $\delta l_{\rm rad} \sim
R_c/\gamma_{\rm eff}$ is sensitive to the bulk Lorentz factor $\bar\gamma_{0,\rm ex}$,
\be\label{eq:Nwave}
N \;\sim\; 
k_z'\cdot\delta l_{\rm rad}' \;\gtrsim\; 10 {(\bar\gamma_0')^2\over \tilde s'};
\ee
Recalling that $\bar\gamma_0' = \bar\gamma_0/(1+\beta_{0,\rm ex})\bar\gamma_{0,\rm ex}$,
this gives $N \gtrsim 3(\bar\gamma_0/\bar\gamma_{0,\rm ex})^2$ when $\bar\gamma_0 > \bar\gamma_{0,\rm ex} \gg 1$.

Our fourth and principal
consideration is the net energy radiated by the charged solitons clustered within the radiation length $\delta l_{\rm rad}$.
These solitons will generally have different sizes and group speeds, and so will interact with each other; this
interaction cannot, of course, be specified using linear theory.  In what follows, we adopt a rough approach
of treating the soliton charge as a Poisson process, meaning that the net charge $Q_N$ carried within the radiation length
is increased to $Q_N^2 \sim N Q^2$.   The energy radiated per unit charged particle is then independent of $N$,
and may be estimated using the formula (\ref{eq:Ecperp}) for an isolated charge.

It is useful to compare the radiated energy with the (unperturbed) kinetic energy of the particles overlapping a soliton.
We start by computing $\delta E_c^\perp$ in Equation (\ref{eq:Ecperp}) in the case of a static medium outside
the current sheet, $\bar\gamma_{0,\rm ex} = 1$:
\be\label{eq:Eradnorm}
         {\delta E_c^\perp\over (k_y^{-1}k_z^{-1}\,2\Delta) n_0\bar\gamma_0 mc^2}  \;\sim\;
         7\times 10^{-3} {\widetilde Q^2\over \bar\gamma_0\,k_{p,\rm ex}\Delta}
         \left({k_yk_z\over k_{p,\rm ex}^2}\right)^{-1} \,\left({\omega_c^{\rm peak\perp}\over ck_{p,\rm ex}}\right)
           \left(\varepsilon_\rho{|J_{z1}|\over J_{z0}}\right)^2.
\ee
Figure \ref{fig:power} shows the result, with the nonlinearity parameter $\varepsilon_\rho |J_{z1}|/J_{z0} \sim 1$
and $\omega_c^{\rm peak\perp}$ and $\widetilde Q$ evaluated using Equations (\ref{eq:omcpk}) and (\ref{eq:tildeQ}).
We also take the transverse size $k_y^{-1}$ of the wavepacket along the current sheet to be comparable to the field penetration depth
${\rm Re}[\kappa_{\rm ex}]^{-1} \sim \bar\gamma_0/k_{p,\rm ex}$ outside the current sheet;
as a result, the flow Lorentz factor $\bar\gamma_0$ scales out of the result.

One finds that $\delta E_c^\perp$ is about $10^{-4}$ of the particle kinetic energy when $k_z$ (equivalently,
the soliton size) is chosen to
maximize the curvature frequency ($k_z \sim 0.7\,k_{p,\rm ex}$;  see Figure \ref{fig:omcpk}).  Note also that the
plasma density scales out of the result when the mode wavevector scales with $k_{p,\rm ex}$.
A larger net energy radiated over a distance $\sim r$, by a factor $\sim \gamma_{\rm eff}r/R_c \sim 10-10^2$.

\section{Implications for Radio Pulsars and Magnetars}\label{s:pulsars}

We now summarize how the radio emission of ordinary (non-Crab-like) pulsars
may arise from the combination of instabilities described here and in Paper I.
Our basic premise is that the polar magnetic field is not a passive actor in the process of radio emission
from ordinary pulsars.  At some level, this should not come as a surprise, given the role
that a dynamic equatorial current sheet may play in the emission of Crab-like pulsars
(e.g. \citealt{cerutti17}).  The processes investigated here are subtler, and we suggest
have not yet been probed by global PIC simulations.
Magnetic fields with inhomogeneous twist distributions are long known to be susceptible to small-scale resistive
instabilities \citep{white13}.  Slow magnetic tearing on scales as small as the plasma skin depth, leading to
cross-field modulations of the bulk plasma motion, is shown in this paper to trigger the Cerenkov emission of charged, subluminal
electromagnetic modes.  The radiative properties of these modes have been investigated.

We have identified a serious candidate for a radio emission process, coherent curvature radiation
of X-mode photons, that operates well below the plasma
cutoff and therefore is consistent with inferences of a high pair multiplicity in the open pulsar circuit.
The particle Lorentz factor leading to efficient linear growth of trapped Alfv\'en modes
($\bar\gamma_0 = 50-100$) is consistent with models of a pair cascade in the pulsar polar cap
(e.g. \citealt{hibschman01,th19}).

At the same time, both instabilities described
here and in Paper I will also operate in parts of the pulsar circuit where pair creation is weak or absent.
It is suggested that sub-pulse drift involves resistive modes that form in pair-free parts of the pulsar
circuit and trigger secondary pair cascades during their nonlinear development.

The severity of the constraints imposed by a high pair multiplicity is worth emphasizing.
Modelling of nebular synchrotron emission surrounding the
Vela pulsar implies a multiplicity ${\cal M}_\pm \sim 10^5$ of positrons relative to seed corotation charges
\citep{dejager07,bucciantini11}.  
Taking appropriate neutron star parameters (polar magnetic field $B_p = 6.8\times 10^{12}$ G,
spin period $P = 0.089$ s) and a mean secondary pair Lorentz factor $\bar\gamma_\pm \sim 10^2$ and
dispersion $\delta\gamma_\pm \sim 10$, one finds
\be
   {\omega_p(r)\over 2\pi} = {\bar\gamma_\pm\over 2\pi}\left[{2{\cal M}_\pm\over\delta\gamma_\pm\bar\gamma_\pm}
       {4\pi|e|B_p\over m_ecP}  \left({r\over r_{\rm NS}}\right)^{-3}\right]^{1/2} \sim
   920~{\rm GHz}\,\left({{\cal M}_{\pm}\over 10^5}\right)^{1/2}
   \left({\bar\gamma_{\pm,2}\over \delta\gamma_{\pm,1}}\right)^{1/2}
     \left({r\over 10~r_{\rm NS}}\right)^{-3/2}.
\ee
Here, $r_{\rm NS}$ is the neutron star radius.
The plasma frequency in the frame of the star is related to the (primed) rest frame of the pairs
by $\omega_p = \bar\gamma_\pm \omega_p' = \bar\gamma_\pm[4\pi e^2(n_+'+n_-')/\delta\gamma_\pm m_e]^{1/2}$,
where $\bar\gamma_\pm(n_+'+n_-') \simeq 2{\cal M}_\pm |\rho_{\rm co}/e|$ and $|\rho_{\rm co}| = \Omega B/2\pi c$
is the corotation charge density \citep{gj69}.
Maser emission of 100 MHz-GHz photons is restricted to the
O mode, and so is ruled out in zones where ${\cal M}_\pm$
is even moderately larger than unity.

\subsection{Low-Frequency Curvature Emission of Extraordinary Mode Photons}

{\it Cross-Field Inhomogeneity.}
The Cerenkov instability of charged Alfv\'en waves is driven by
a gradient in the particle flow speed, on a scale comparable to the
plasma skin depth $k_{p,\rm ex}^{-1} \sim 0.1 - 100$ cm,
with this gradient oriented perpendicular to the magnetic field.
This instability does not depend on any irregularity in the particle distribution as measured
on a single magnetic flux element.   A gradient in the distribution function across field lines
can be sustained for a long time compared with the plasma timescale $(ck_{p,\rm ex})^{-1}$.

{\it Charge Clumping.}  
Soliton formation depends on the fast growth of trapped Alfv\'en modes, which in turn
implies particle Lorentz factors below $\bar\gamma_0 \sim 50-10^2$ (Equation (\ref{eq:gam0max})
and Figure \ref{fig:omcpk}).
An Alfv\'en soliton size
$k_z^{-1} \sim k_{p,\rm ex}^{-1}$ is inferred from the $k_z$-dependent of the linear growth rate.
 A detailed exploration of the formation of Alfv\'en solitons would most productively involve
  PIC simulations of the linear instability.

{\it X-mode Curvature Emission below the Ambient Plasma Frequency.}
The requirement of rapid mode growth also forces the peak curvature frequency well below the ambient plasma cutoff,
to $\omega_c^{\rm peak\perp} \sim 10^{-2}\,ck_{p,\rm ex}$
(Figure \ref{fig:omcpk} and Equation (\ref{eq:omcpk3})).  As a result, the emitted radiation is concentrated
entirely in the X mode, with electric vector oriented perpendicular to the background magnetic field.
It has been argued that the core pulsed radio emission of Vela and several 
other pulsars is concentrated in the X mode \citep{rankin15}.

It should be emphasized that the kinetic approach adopted here incorporates the effects of Debye screening on the
trapped Alfv\'en modes.  The transverse field and current perturbations are found to
penetrate a substantial distance $\sim \bar\gamma_0k_{p,\rm ex}^{-1}$ into the medium outside the supporting
current sheet.  On this basis, we suggest that the strong screening that has been claimed for low-frequency
X-mode emission by electrostatically driven charge clumps \citep{glm04} does not apply to Alfv\'en solitons.

A suppression of the peak emission frequency far below $\omega_{p,\rm ex} = ck_{p,\rm ex}$
is consistent with the emission of 100 MHz-GHz photons if the pair multiplicity
is as high as suggested by some modelling of nebular synchrotron radiation (${\cal M}_\pm \sim 10^5$;
\citealt{dejager07,bucciantini11}.)  Emission in this band is then inconsistent with
any mechanism producing ordinary (O-mode) waves near the plasma cutoff.
Detections of pulsars by LOFAR \citep{bilous20, bondonneau20} are especially challenging in this respect.

The detection of orthogonal polarization components in some radio pulsars (e.g. \citealt{stinebring84})
is not clearly inconsistent with uniform emission in a single polarization mode.  Propagation near
the cyclotron resonance and the polarization-limiting surface can leave a significant imprint on the
polarization (e.g. \citealt{petrova00,beskin12}), as can a more complicated surface magnetic field
(e.g., involving strong high-order multipoles).

{\it Emitted Radio Power.}  Only a modest fraction ($\sim 10^{-4}$)
of the particle kinetic energy flowing through the current-carrying magnetosphere
can be converted to curvature photons over the radiation decoupling length $\delta l_{\rm rad} =
R_c/\gamma_{\rm eff}$
(Figure \ref{fig:power} and Equation (\ref{eq:power2})).  Saturation of this bound requires that 
trapped Alfv\'en wavepackets absorb a significant fraction of the kinetic energy of the charge flow.

The net flux of kinetic energy carried by secondary pairs 
away from a pulsar of spin period $P = 2\pi/\Omega$ and polar magnetic field $B_p$ is
\be
L_{\rm kin} \simeq 2{\cal M}_\pm{\Omega^2B_pR^3\over 2c|e|}{\bar\gamma_0 m_ec^2}
= 2\times 10^{28}\,{\cal M}_\pm\,P_{-1}^{-2}B_{p,12}R_6^3\bar\gamma_{0,2}\quad{\rm erg~s^{-1}}.
\ee
By way of comparison, the observed radio output of pulsars shows a large dispersion but remarkably
little apparent variation with spindown power, mostly ranging between
$\sim 10^{28}$ and $10^{30}$ erg s$^{-1}$ \citep{szary14}.  Taking a radiation efficiency of $10^{-3}$
(accumulated over the full radius $r \sim 10^{1-2}\,\delta l_{\rm rad}$)
one sees that a pair multiplicity ${\cal M}_\pm \sim 10^4\,P_{-1}^2B_{12}^{-1}R_6^{-3}L_{\rm rad,29}$
is required to generate a radio luminosity $L_{\rm rad}$.

To summarize:  we infer a large pair multiplicity in the open pulsar circuit,
so as to compensate the low curvature frequency and emission efficiency of Alfv\'en solitons.

{\it Radius-to-frequency mapping.}
The underlying resistive instability (as described in Paper I) operates in both pair-rich and pair-starved
portions of the current-carrying magnetosphere.  It will be triggered at high and low altitudes above
the neutron star surface.  The characteristic current sheet width ($\Delta \sim 1/k_{p,\rm ex}$) is downscaled
by a factor $\sim {\cal M}_\pm^{-1/2}$ in the presence of a high abundance of pairs, but the
growth rate of the tearing instability is not affected.  The instability was studied in Paper I for $k_z = 0$,
corresponding to a great extension of the current sheets formed along the mean helical magnetic field.
Furthermore, the efficiency of curvature radiation by Alfv\'en solitons is found not 
to depend on the particle density and therefore on
the size of the charge clumps. The basic radial scaling $\omega_c^{\rm peak\perp} \propto
\omega_{p,\rm ex} \propto B^{1/2} \propto r^{-3/2}$ will be maintained
at large altitudes, where the pair multiplicity
has saturated;  but strong deviations from this scaling are expected within the electromagnetic cascade,
where ${\cal M}_\pm$ grows rapidly.

\subsection{Maser Emission?}\label{s:maser}

Let us now consider whether the interaction of the primary $e^\pm$ beam with a
Cerenkov-unstable trapped
Alfv\'en wave can stimulate the emission of a secondary electromagnetic wave, to be identified
ith the pulsar radio wave.  Viewed classically, this process involves the excitation of a secondary charge-density wave
and so requires that the Alfv\'en wave be narrow band \citep{lyubarskii96}.

The fastest-growing trapped Alfv\'en mode is described in Appendix \ref{s:peak}.
It is of interest for maser emission because (i) it is localized
in a narrow range of $k_z$ and (ii) represents a maximum in the longitudinal energy of the mode relative to
the transverse wave energy.   The importance of the longitudinal electric field for maser growth
is two-fold.  First, the transverse displacement of the pairs in the Alfv\'en wave is
dominated by ${\bf E}\times{\bf B}$ drift,
and so is suppressed relative to the longitudinal displacement by the ratio $\sim \phi_1/\Delta B_{z0}$
of the transverse electric field to the guide magnetic field.  Indeed, \cite{fung04} find, in their
numerical experiment of maser action in a current free Alfv\'en wave ($k_\perp = 0$), that charged particle
bunching turns off as the guide field rises above the wave field.  
Second, when the pair multiplicity ${\cal M}_\pm$ is high, the ${\bf E}\times{\bf B}$ drift rates
of positive and negative charges nearly cancel.  By contrast, neither of these effects suppresses
the interaction of the primary $e^\pm$ flow with the parallel wave electric field $E_{\parallel,1}$ --
as simulated by \cite{schopper03} in the case of a Langmuir wave pump.

Focusing now on the ``longitudinal maser'' channel, one notes that it has the potential to provide
a second polarization component, orthogonal to the emitted radio wave, as is observed in some pulsars
\citep{stinebring84}.  Nonetheless, other questions arise
about its viability when seeded by a trapped Alfv\'en mode.

First, Alfv\'en waves near peak growth show a broader range of phase speed
than of $k_z$ (see Figure \ref{fig:numdisp} in Appendix \ref{s:peak}).
The electromagnetic wave frequency $\omega_2 \simeq c|{\bf k}_2|$ is related
to the Alfv\'en mode frequency $\omega$ by $\omega_2-(\bar\bbeta_0c)\cdot{\bf k}_2 =
\omega - \bar\beta_0 ck_z = \bar\omega$ giving $|\omega_2| \simeq
2\bar\gamma_0^2|\omega-\bar\beta_0 ck_z| = 2ck_{p,\rm ex}|\varpi|$ (see Equation (\ref{eq:dimensionless})).
It is therefore questionable
whether the resonance condition between the Alfv\'en mode and the scattered electromagnetic wave could be
satisfied for a realistic range of excited Alfv\'en modes.  

Second, maser photons seeded by an Alfv\'en mode of peak growth would have an uncomfortably
high frequency, $|\omega_2| \sim 0.56\bar\gamma_0^{1/2}\omega_{p,\rm ex} \gg \omega_{p,\rm ex}$,
well beyond the plasma cutoff (see Equation (\ref{eq:speak})).   This constraint could be relaxed
if the radio wave were excited by the interaction of the seed particle flow with fully developed
Alfv\'en solitons that propagate at a ratio $\gamma_{\rm gr}/\bar\gamma_0$ closer to unity.  Then
the escaping wave could have a frequency $|\omega_2| \sim \omega_{p,\rm ex}$.

\subsection{Fast and Slow Radio Flux Variations}

Pulsar radio emission varies over a combination of fast and slow timescales, as normalized by
the stellar rotation \citep{gs03}.
The characteristic timescale for particles and electric field to oscillate in a
near-surface gap of height $h_{\rm gap} \sim 0.1-1$ km
is $t_{\rm gap} \sim h_{\rm gap}/c \sim 0.3-3\,\mu$s.  This is several 
orders of magnitude shorter than the rotation period, and is far too short to explain the strong stochastic variations
in pulse shape that are frequently seen over successive rotations.

The resistive instability described in Paper I provides a quantitative starting point
for investigating relatively slow phenomena such as sub-pulse drift.   The Cerenkov instability described in this paper,
along with the associated curvature emission, is modulated by the slow resistive instability.

{\it Slow, Stochastic Variations in Pulse Structure.}  Magnetic tearing slowly feeds off high-order variations
in the current in the open pulsar circuit, the growth rate being suppressed
by the inverse of the strong poloidal magnetic field, $s_{\rm tear} \gtrsim 4\pi (k_y/k_x) |J_z|/B_z$
(Paper I).  Here, $B_z$ is the strong poloidal (`guide') magnetic field,
$k_y$ is the wavevector component in the direction perpendicular to ${\bf B}$ and
parallel to the current sheet, and $k_x$ is the characteristic gradient scale of the magnetic field transverse to the sheet.  Taking the small-scale current density to be comparable in magnitude to the mean current
driven by the corotation charge flow, $J_z \sim \rho_{\rm co}c$, gives $s_{\rm tear} \gtrsim 2\Omega (k_y/k_x)$.
Strong pulse-to-pulse variations in the pattern of magnetic tearing in the open-field bundle of
a pulsar are a natural consequence of this slow growth.

{\it Microstructure.}  In the approach described here, the time profile of a single radio pulse depends
on the spatial distribution of compact tearing surfaces.  Curvature radiation by Alfv\'en solitons is
beamed and localized on small-scale current sheets
of a thickness $\Delta \sim k_{p,\rm ex}^{-1} \sim 0.1-100$ cm (depending on $r$ and ${\cal M}_\pm$).  When this
beamed emission is swept past the line of sight of the observer by the pulsar's rotation, it
can form sub-pulse structure on a timescale as short as
\be
t_{\rm micro} \sim {1\over\Omega}{\rm max}\left[{1\over\gamma_{\rm eff}}, {\Delta\over r}\right].
\ee
The first term on the right-hand side represents the angular broadening as limited by the bulk relativistic
motion of the charge clumps, and the second the finite angle subtended by the current sheet.
The first term easily dominates when $\Delta \sim 1/k_{p,\rm ex}$.
Curvature emission at a frequency $\omega_c^{\rm peak\perp}/2\pi \sim 1$ GHz and an altitude
of 100 km is beamed into an angle $1/\gamma_{\rm eff} \sim 0.005$, meaning that
$t_{\rm micro} \sim 10^{-4} (P/1~{\rm s})$ s.

More generally, the imprint of current perturbations of transverse size
$\Delta < r/\gamma_{\rm eff} \sim 0.5~r_7$ km will be washed out by relativistic beaming.
A model determining the spectrum of current perturbations
in a pulsar magnetosphere is beyond the scope of this paper.

\subsection{Sub-pulse Structure and Drift}

The detection of sub-pulse drift in many pulsars (e.g. \citealt{deshpande01}) suggests
the presence of electromagnetic field structures that persist over multiple rotations.
Indeed, the phase of sub-pulses has a remarkable ability to survive periods of nulling \citep{fr82}.
Although a rough description of sub-pulse drift in terms of ${\bf  E}\times{\bf B}$ drift has long been
suggested \citep{rs75}, no concrete explanation has been offered as to how a stochastic longitudinal
excitation of the pairs, with a characteristic timescale of $t_{\rm gap} \sim 1\,\mu$s, could
maintain phase coherence over a period of seconds.

By contrast, the structures created by a tearing instability generically have a lifetime longer than the
rotation period and experience angular drift (Paper I). 

{\it Sub-pulse drift.}  In a charge-asymmetric plasma, the tearing instability 
is an overstability with a finite real frequency,
\be
\omega_{\rm tear} \sim 2\Omega {k_y\over k_x}{|n_0^+-n_0^-|\over n_0^++n_0^-},
\ee
where $n_0^\pm$ are the space densities of positive and negative electrons (see Figure 8 of Paper I).
The tearing structure has a phase speed parallel to the current sheet,
\be
v_y = {\omega_{\rm tear}\over k_y} \sim 
{c\bar\gamma_0^{1/2}\over \widetilde k_x\,(1+2{\cal M}_\pm)^{3/2}}\left({2\Omega\over |e|B_z/m_ec}\right)^{1/2},
\ee
in the case where the corotation charge density is negative and the primary charge flow is provided
by electrons.  Here, we have normalized $k_x^2 = \widetilde k_x^2 \cdot 4\pi e^2(n_0^+ + n_0^-)/\bar\gamma_0m_ec^2$.

This drift can operate either in a prograde or retrograde sense with respect to the direction of rotation,
depending on the sign of $k_y$.  
The drift rate is suppressed by a high pair multiplicity, implying that the dominant drifting structures
are those forming in parts of the pulsar polar cap that experience weak (or absent) pair creation
in the absence of tearing.  Focusing still on the case of negative corotation charge, we
normalize the electron Lorentz factor by the voltage drop across the open magnetic field lines,
$\bar\gamma_0 = \varepsilon_\gamma |e|\Phi_{\rm open}/m_ec^2$, where
$\Phi_{\rm open} = {1\over 2}(\Omega r_{\rm NS}/c)^2 B_p r_{\rm NS}$ and generally $\varepsilon_\gamma \ll 1$.
The time to rotate through a radian in azimuthal angle at a perpendicular distance $R = (\Omega r/c)^{1/2}r$ from the
magnetic axis is, simply,
\be\label{eq:trot}
t_{\rm rot} \;\sim\; {R\over v_y} \;\sim\;  {\widetilde k_x\over \varepsilon_\gamma^{1/2}\Omega}.
\ee
This result is similar to that originally obtained by \cite{rs75} in their heuristic model of ${\bf E}\times{\bf B}$
drift of charge clouds in the pulsar polar cap.

We are led to the following interplay between tearing, pair creation, and radio emission.
The drift timescale (\ref{eq:trot}) is similar to that observed when the $e^\pm$ are very relativistic (but
still below the threshold for emitting curvature gamma rays that trigger a pair cascade).
The current density increases
within the drifting structure over the background, thereby potentially igniting pair creation in parts
of the structure.  The energy of the secondary $e^\pm$ pairs is in the range where current sheets experience rapid
Cerenkov-driven growth of charge density waves.  Because the drift is slowed by $e^\pm$ creation
(by a factor $\sim \Delta r\Omega/c$, where $\Delta r$ is the height of the pair creation front),
the zones experiencing a cascade must comprise a relatively small part of the current structure
for its drift to continue.

{\it Sub-pulse `Carousels'} (e.g. \citealt{deshpande01}).
An azimuthally periodic array of magnetic O- and X-points is seen in numerical calculations of
high-order magnetic tearing near a local extremum in an axially symmetric twist profile
(e.g. \citealt{bierwage05}).
Analogous high-order tearing modes were demonstrated in Paper I in cartesian symmetry.
Azimuthally periodic structures forming in the pulsar polar cap
would drift over the timescale given in Equation (\ref{eq:trot}).  The quoted global simulations
of high-order magnetic tearing have the interesting implication
that coordinated sub-pulse drift arises more easily in pulsars with more ordered, dipolar magnetic
fields, and also with smaller inclinations of the dipole axis with respect to the axis of rotation.

\subsection{Implications for Magnetars}

Some further implications of our results for magnetar electrodynamics will be briefly summarized here.
The strongly stochastic, pulsed radio emission that is observed from a handful of magnetars provides
perhaps the most direct evidence for a role for magnetic tearing in persistent radio emission.

Emission by the curvature process described here depends on the presence of
relativistic $e^\pm$ in order to sustain trapped Alfv\'en waves with a relativistic group speed.
Observations of persistent hard X-ray emission from magnetars
point to dissipation rates exceeding $10^3$ times the spindown power of the neutron star
in some cases \citep{kb17}.
The plasma state of the magnetosphere has been debated:  alternative models include (i) a relativistic
double layer supported by counterstreaming $e^-$ and $e^+$ with $\bar\gamma_0 \sim 10^3$
\citep{bt07} and (ii) a trans-relativistic,
collisional plasma that is sustained by intense ohmic heating in localized dissipative structures \citep{tk20}.
The second option more directly accounts for the hard X-ray spectra
through soft-photon emission associated with $e^+-e^+$ annihilation.  The dissipative structures naturally
arise from fault-like features in the magnetar crust \citep{tyo17}, and they are a source of
gamma rays of energy $\sim m_ec^2$.    Nonlocal collisions of these gamma rays, $\gamma + \gamma \rightarrow e^+ + e^-$,
are expected to fill much of the remainder of the magnetosphere with transrelativistic pairs
that are inefficient sources of coherent curvature radiation.

It should be kept in mind that the magnetic field in the outermost magnetosphere, and the open circuit,
must be strongly dynamic when a magnetar is in active state,
especially if the transport of magnetic twist is mediated significantly by internal tearing \citep{t08}
in addition to ohmic diffusion \citep{beloborodov09}.  Internal tearing provides a promising framework
for understanding
the extreme stochasticity of the pulsed radio emission of some magnetars \citep{levin12,yan15,dai19}.
We have previously argued that the current flowing
through the outer magnetosphere will relax to a metastable configuration, with the outward transport
of twist being balanced by reconnection near the magnetospheric boundary.  The strong secular changes in pulse
profile that are observed in radio magnetars \citep{camilo07} plausibly
reflect secular changes in the current profile driven by internal tearing.  The high peak frequency
($O(100)$ GHz:  e.g. \citealt{chu21}) points to a high current density and a cascade process that drives
small-scale braiding of the magnetic field.

\acknowledgements
We acknowledge the support of the Natural Sciences and Engineering Research Council of Canada (NSERC) through grant RGPIN-2017-06519.

\appendix

\section{Peak Growth Rate}\label{s:peak}

The linear growth rate of a trapped Alfv\'en mode is causally limited when the particle flow in the current sheet
is relativistic.  We now identify the mode with maximum growth rate $s_{\rm max}$.  It should be emphasized
that the fastest growing modes are not those of greatest interest for radio emission, because the group
speed of the mode drops substantially as $k_z \rightarrow k_{p,\rm ex}$ (Figure \ref{fig:velocity}).  
Charge packets whose curvature emission peaks at a frequency $\omega_{p,\rm ex} = ck_{p,\rm ex}$ grow most rapidly if
their longitudinal wavenumber $k_z$ is slightly offset below $k_{p,\rm ex}$, and $k_{p,\rm ex}\Delta \sim 1$.  Packet
growth as constrained by curvature radiation efficiency is discussed in Section \ref{s:power}.

To obtain $s_{\rm max}$, we must include terms of higher order in $\bar\omega/ck_z$, $s/ck_z$
in the eigenvalue equation as compared with the truncated Equation (\ref{eq:disp2}).  In effect, we must include the
full expression for $J_{z1}$ within the
currrent sheet.  Then Equation (\ref{eq:disp}) with the variable substitutions (\ref{eq:dimensionless}) becomes
\be\label{eq:disp4}
\left[{k_z^2\over k_{p,\rm ex}^2} - 1 + 2{\varpi + i\tilde s\over\bar\gamma_0^2}\right](\varpi + i\tilde s)^4 =
(1 - 2\varpi - 2i\tilde s)\left[1 - {(\varpi + i\tilde s)^2\over\bar\gamma_0}\right]^2(k_{p,\rm ex}\Delta)^2.
\ee
We will solve this equation in two regimes, both satisfying the inequalities $k_{\rm p,\rm ex}\Delta \gg 1/\bar\gamma_0$
and $|\varpi + i\tilde s| \gg 1$.

When $k_{p,\rm ex}\Delta < 1$, peak growth corresponds to setting $k_z^2/k_{p,\rm ex}^2 - 1 \rightarrow 0$ in
Equation (\ref{eq:disp4}), and we obtain an equation for the variable $Y = (\varpi + i\tilde s)/\bar\gamma_0^{1/2}$,
\be
{Y^2\over 1-Y^2} = e^{i(2n+1)\pi/2} k_{p,\rm ex}\Delta\quad\quad (n \in \mathbb{Z}).
\ee
The solution consistent with the constraint ${\rm Re}[\kappa_{\rm in}^2] < 0$ is
\be
s_{\rm max} = \left({k_{p,\rm ex}\Delta\over 2}\right)^{1/2}{ck_{p,\rm ex}\over\bar\gamma_0^{3/2}}\quad\quad
(\bar\gamma_0^{-1} \ll k_{p,\rm ex}\Delta \ll 1).
\ee
Notice that the dependence of $s_{\rm max}$ on $\bar\gamma_0$ has softened to 
$\bar\gamma_0^{-3/2}$.  This also means that the group Lorentz factor of the fastest-growing mode increases more slowly than the
particle Lorentz factor, $\gamma_{\rm gr} \propto \bar\gamma_0^{3/4}$.

\begin{figure}
  \epsscale{0.55}
  \vskip -0.3in
\plotone{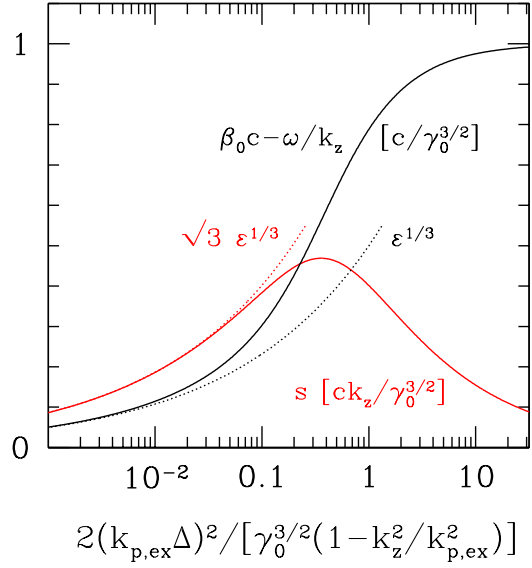}
  \vskip -.5in
  \caption{Numerical solution to Equation (\ref{eq:disp5})
    for the mode growth rate $s = \tilde s\, ck_z/\bar\gamma_0^2 \propto \bar\gamma_0^{-3/2}$ (red solid curve) and
    group speed lag
    $\bar\beta_0c - \omega/k_z$ (black solid curve).  Dotted curves show the scaling solution for $|\varepsilon| > 1$
    given by Equation (\ref{eq:mode0}).\label{fig:numdisp}} 
\end{figure}

In the second regime where $k_{p,\rm ex}\Delta > 1$, peak growth occurs at small but finite $k_z^2/k_{p,\rm ex}^2 - 1$,
and we must instead solve
\be\label{eq:disp5}
{Y^3\over (1-Y^2)^2} \;=\; {2(k_{p,\rm ex}\Delta)^2\over\bar\gamma_0^{3/2}(1-k_z^2/k_{p,\rm ex}^2)} \;=\;
{8|\varepsilon|\over\bar\gamma_0^{3/2}}
\ee
for a range of $k_z/k_{p,\rm ex}$ and then find the maximum growth rate.
This equation has one root with negative $\varpi$ and positive $\tilde s$.  The numerical solution is shown in Figure \ref{fig:numdisp}.
The peak growth rate is
\be\label{eq:speak}
s_{\rm max} \simeq 0.47\, {ck_{p,\rm ex}\over\bar\gamma_0^{3/2}};\quad\quad \varpi \simeq -0.56 {ck_{p,\rm ex}\over\bar\gamma_0^{3/2}}
\quad\quad (k_{p,\rm ex}\Delta \gg 1),
\ee  
which is attained at $1-k_z^2/k_{p,\rm ex}^2 \simeq 5.6\,(k_{p,\rm ex}\Delta)^2/\bar\gamma_0^{3/2}$ and
$|\varepsilon| \simeq 0.045\bar\gamma_0^{3/2}$.  Figure \ref{fig:numdisp} shows that the scaling solution
given by Equation (\ref{eq:mode0}) is accurate at lower $|\varepsilon|$ and smaller growth rates.  Once again, the group Lorentz factor $\gamma_{\rm gr} \propto \bar\gamma_0^{3/4}$.

The fastest-growing mode is most strongly localized about the current sheet at $k_{p,\rm ex}\Delta > 1$.
Equation (\ref{eq:kapin}) generalizes to
\be
\kappa_{\rm ex}^2 = (1-2\varpi-2i\tilde s)\left[{k_z^2-k_{p,\rm ex}^2\over\bar\gamma_0^2} +
  {2(\varpi + i\tilde s)k_z^2\over\bar\gamma_0^4}\right].
\ee
This gives
\be
   {\rm Re}[\kappa_{\rm ex}] \;\simeq\; {k_{p,\rm ex}\over\bar\gamma_0^{3/2}}\times
   \begin{cases}  (2k_{p,\rm ex}\Delta)^{1/2}\quad\quad & (k_{p,\rm ex}\Delta < 1); \\
     0.58\,k_{p,\rm ex}\Delta \quad\quad & (k_{p,\rm ex}\Delta > 1).
     \end{cases}
\ee

\section{Effects of Particle Streaming Outside the Current Sheet}\label{s:streamext2}

Most results in the main text are derived for the simplest case where particle streaming is
confined to narrow current sheets.
Given the small sheet thickness, $\Delta \sim 1/k_{p,\rm ex}$, weaker relativistic streaming may
also be present outside a given sheet, especially in the situation where most of the charges are the end product
of an electromagnetic cascade in the pulsar polar cap.   As before, the particle density is assumed to be uniform
everywhere, but now the medium at $|x| > \Delta$ moves with bulk Lorentz factor $\bar\gamma_{0,\rm ex}$
in the frame of the star.
As explained in Section \ref{s:streamext}, the mode growth rate $s$ can be obtained by
a local Lorentz boost along $\hat B$ to the rest frame outside the sheet.  In this frame, henceforth labelled
by a prime ($'$), we may deploy the previously derived result, now evaluated in terms of $k_z'$ and $k'_{p,\rm ex}$.The transverse wavenumber $k_y$ and the sheet thickness $\Delta$ are invariant under the boost.

The main qualitative effects are (i) to increase the peak curvature frequency that a rapidly growing
wave may radiate and (ii) to boost the net power of the radiation.
As a wavepacket propagates a distance $\delta l = \int \hat B\cdot {\bf dl}$ along a current sheet
that is aligned with the polar magnetic field of a neutron star, it experiences a lapsed time
$\delta t' = \bar\gamma_{0,\rm ex}(1-\bar\beta_0\bar\beta_{0,\rm ex})\,\delta l/\bar\beta_0c \simeq
\bar\gamma_{0,\rm ex}(1-\bar\beta_{0,\rm ex})\delta l/c$. The discussion that follows is simplified by
assuming that $\bar\gamma_{0,\rm ex} \ll \bar\gamma_0$, which allows us to set factors of $\bar\beta_0$ to unity.

We first re-evaluate the curvature frequency in the frame of the star.
Choosing the same growth criterion, $s' \gtrsim 10/\delta t'$, and choosing $\delta l = \delta l_{\rm rad} = R_c/\gamma_{\rm gr}$,
where $\gamma_{\rm gr} \simeq \bar\gamma_{0,\rm ex}(1+\bar\beta_{0,\rm ex})\gamma_{\rm gr}'$ is the
group speed in the frame of the star, we get
\be\label{eq:omcpk3}
   {\omega_c^{\rm peak\perp}\over ck_{p,\rm ex}} =
    (1+\bar\beta_{0,\rm ex})
   \left({\omega_c^{\rm peak\perp}\over ck_{p,\rm ex}}\right)_{\bar\gamma_{0,\rm ex} = 1}.
\ee
Here, as before,
$ck_{p,\rm ex} = \bar\gamma_{0,\rm ex}ck_{p,\rm ex}'$ is the Lorentz-boosted cutoff frequency.
One finds only a modest change in peak curvature frequency compared with the case of static plasma
outside the current sheet.

Bulk streaming outside the current sheet introduces only a modest adjustment on the upper bound on $\bar\gamma_0$
for fast linear Alfv\'en mode growth (Equation (\ref{eq:gam0max})),
\be\label{eq:gam0max2}
\bar\gamma_0 < 
(1+\beta_{0\rm ex})^{1/3}
\left({\bar\gamma_0'\over\gamma_{\rm gr}'}{k_z'\over k_{p,\rm ex}'}{\tilde s'\over 10}\right)^{1/3}\,
(k_{p,\rm ex}R_c)^{1/3}.
\ee
Here, we have made use of the rescaling of $s'$ in Equation (\ref{eq:srest}).

There are also implications for the efficiency of curvature radiation if 
$\bar\gamma_{0,\rm ex}$ is not much smaller than $\bar\gamma_0$, meaning that $\bar\gamma_0'$ is not much
larger than unity (Section \ref{s:power}).
The charge carried by a rapidly growing mode is invariant under the boost $\bar\gamma_{0,\rm ex}$ and
the radiated energy moderately enhanced,
\be\label{eq:power2}
Q \;=\; Q\bigr|_{\bar\gamma_{0,\rm ex}=1}; \quad\quad
{\delta E_c^\perp\over (k_y^{-1}k_z^{-1}\,2\Delta) n_0\bar\gamma_0 mc^2}  \;=\;
(1+\bar\beta_{0,\rm ex})\bar\gamma_{0,\rm ex}\,
\left[{\delta E_c^\perp\over (k_y^{-1}k_z^{-1}\,2\Delta) n_0\bar\gamma_0 mc^2}\right]_{\bar\gamma_{0,\rm ex} = 1}.
\ee
Here, the nonlinearity parameter $\varepsilon_\rho |J_{z1}'|/J_{z0}'$ has been held constant and evaluated in the primed frame. 
Given a close packing of the clumps,
we end up with a net enhancement of the radio power by a factor $\sim 2\bar\gamma_{0,\rm ex}$, which would 
allow the radiated X-mode energy to reach $\sim 10^{-3}$ of the particle kinetic energy.

\end{document}